\documentstyle[prc,twocolumn,aps,epsf,eqsecnum,ifthen,epsfig]{revtex}

\renewcommand{\P}{\partial}
\renewcommand{\phi}{\varphi}

\newcommand{\nc}{\newcommand}
\nc{\pt}{p_{\rm T}}
\nc{\se}{\section}
\nc{\suse}{\subsection}
\nc{\beq}[1]{\begin{equation}\label{#1}}
\nc{\eeq}{\end{equation}}
\nc{\bea}[1]{\begin{eqnarray}\label{#1}}
\nc{\eea}{\end{eqnarray}}
\nc{\bce}{\begin{center}}
\nc{\ece}{\end{center}}
\nc{\bit}{\begin{itemize}}
\nc{\eit}{\end{itemize}}
\nc{\bmp}{\begin{minipage}}
\nc{\emp}{\end{minipage}}

\nc{\la}{\langle}       
\nc{\lla}{\left \langle}
\nc{\ra}{\rangle}       
\nc{\rra}{\right \rangle}

\newcommand{\lda}{\langle\!\langle}       
\newcommand{\llda}{\left\langle\!\left\langle}
\newcommand{\rda}{\rangle\!\rangle}       
\newcommand{\rrda}{\right\rangle\!\right\rangle}

\newcommand{\bb}{{\bbox{b}}}

\newcommand{\half}{{\textstyle {1\over 2}}}

\newcommand{\ibid}[3]{{\it ibid.} {\bf #1}, #3 (#2)}

\newcommand{\annrev}[3]{Ann. Rev. Nucl. Part. Sci. {\bf #1}, #3 (#2)}

\newcommand{\prevc}[3]{Phys. Rev. C {\bf #1}, #3 (#2)}
\newcommand{\prevd}[3]{Phys. Rev. D {\bf #1}, #3 (#2)}
\newcommand{\prevl}[3]{Phys. Rev. Lett.\ {\bf #1}, #3 (#2)}

\newcommand{\zpc}[3]{Z. Phys. C {\bf #1}, #3 (#2)}
\newcommand{\jpg}[3]{J. Phys. G {\bf #1}, #3 (#2)}
\newcommand{\epjc}[3]{Eur. Phys. J. C {\bf #1}, #3 (#2)}
\newcommand{\plb}[3]{Phys. Lett. B {\bf #1}, #3 (#2)}
\newcommand{\npa}[3]{Nucl. Phys. {\bf A#1}, #3 (#2)}

\newcommand{\hip}[3]{Heavy Ion Physics {\bf #1}, #3 (#2)}
 
\newcommand{\sjnp}[3]{Sov. J. Nucl. Phys. {\bf #1}, #3 (#2)}

\newcommand{\jcomp}[3]{J. Comp. Phys. {\bf #1}, #3 (#2)}

\preprint{\\CERN-TH/2000-161\\hep-ph/0006129}

\title{Anisotropic transverse flow and the quark-hadron phase transition}   
\author{Peter F.~Kolb$^{a,b}$ \and Josef Sollfrank$^b$, 
        and Ulrich~Heinz$^{a,}$\thanks{On 
        leave of absence from Institut f\"ur Theoretische Physik, 
        Universit\"at Regensburg. Email: Ulrich.Heinz@cern.ch}}
\address{$^a$Theoretical Physics Division, CERN, CH-1211 Geneva 23, 
         Switzerland\\
         $^b$Institut f\"ur Theoretische Physik, Universit\"at
         Regensburg, D-93040 Regensburg, Germany
         }
\date{\today}

\begin{document}

\maketitle


\begin{abstract}

We use (3+1)-dimensional hydrodynamics with exact longitudinal 
boost-invariance to study the influence of collision centrality 
and initial energy density on the transverse flow pattern and 
the angular distributions of particles emitted near midrapidity in 
ultrarelativistic heavy-ion collisions. We concentrate on radial flow 
and the elliptic flow coefficient $v_2$ as functions of the impact 
parameter and of the collision energy. We demonstrate that the finally
observed elliptic flow is established earlier in the collision than the
observed radial flow and thus probes the equation of state at higher 
energy densities. We point out that a phase transition from hadronic 
matter to a color-deconfined quark-gluon plasma leads to non-monotonic 
behaviour in both beam energy and impact parameter dependences which, 
if observed, can be used to identify such a phase transition. Our 
calculations span collision energies from the Brookhaven AGS (Alternating 
Gradient Synchrotron) to beyond the LHC (Large Hadron Collider); the 
QGP phase transition signature is predicted between the lowest available 
SPS (CERN Super Proton Synchrotron) and the highest RHIC (Brookhaven 
Relativistic Heavy Ion Collider) energies. To optimize the chances for 
applicability of hydrodynamics we suggest to study the excitation 
function of flow anisotropies in central uranium-uranium collisions
in the side-on-side collision geometry.

\end{abstract}

\se{Introduction}
\label{sec1}

At a given beam energy, the highest energy densities can be reached
in central collisions (impact para\-me\-ter $b{\,=\,}0$) between the 
largest available nuclei. Hence for many years the exper\-i\-mental 
and theoretical attention has focussed on such collisions. Non-central
($b\ne 0$) collisions are, however, interesting in their own right
since they exhibit new phenomena which are forbidden by azi\-muthal 
symmetry in central collisions between spherical nuclei. For 
non-central collisions the directions of the beam axis and the 
impact parameter $\bb$ define the collision plane, and many interesting
physical phenomena are now non-trivial functions of the azimuthal angle
$\phi$ relative to the collision plane. These include in particular the 
transverse geo\-me\-try of the collision fireball as measured with two-particle
Bose-Einstein correlations (see e.g. \cite{LHW00} and references therein)
and momentum-space anisotropies in the transverse plane due to anisotropic
transverse flow of the fireball matter \cite{flowrev}.

Aside from changing the collision energy, limited variations of the 
energy density of the reaction zone are also possible by varying
the collision centrality. Variation of the initial energy density 
provides the handle for studying phase transitions in nuclear matter, 
in particular the quark-hadron transition at a critical energy density 
$e_{\rm c}{\,\lesssim\,}1$\,GeV/fm$^3$ \cite{lattice}. Non-central 
collisions between spherical nuclei and/or central collisions between 
deformed nuclei provide new opportunities to correlate phenomena related 
to azimuthal anisotropies with the initial energy density. This may 
yield novel phase transition signatures. In \cite{Rischke95} this idea 
was exploited for the so-called directed flow at forward and backward 
rapidities: the softening of the equation of state (EOS) in the phase 
transition region was predicted to lead to a reduction of the directed 
flow, making the phase transition visible as a minimum in its excitation 
function. Sorge \cite{S97,S99} suggested analogous features for the 
elliptic flow \cite{St82,O92,O98} which were further studied in 
\cite{HL99,KSH99,TS99}. The effects of a phase transition on the 
excitation function of radial flow in central collisions between 
spherical nuclei had been discussed earlier in \cite{SZ78,vH82,KRLG86}.

An important difference between the radial flow observed in azimuthally 
symmetric central collisions and the anisotropic directed and elliptic 
flows in non-central collisions and/or central collisions between 
deformed nuclei was pointed out by Sorge in \cite{S97}: 

1. {\em Directed flow} affects mostly particles at forward and backward 
rapidities which (at energies above a few hundred MeV/nucleon) are 
deflected away from the beam direction by the pressure built up between 
the colliding nuclei during the time of their mutual overlap. Since the 
thus affected particles quickly leave the central region where this 
transverse pressure force acts, the finally observed directed transverse 
flow pattern is established very early in the collision. Its natural 
time scale is given by the transition time of the two colliding nuclei 
which decreases with increasing beam energy; this causes a decrease at 
high collision energies (after an initial rise at low beam energies) of 
the directed flow \cite{flowrev}. This decrease is amplified by a lack of
thermalization during the very earliest stages of the collision which
prohibits fast enough buildup of transverse pressure and thus eventually 
invalidates the applicability of hydrodynamic concepts for calculating
the directed flow. Such pre-equilibrium features may even cover up 
\cite{S97} the phase transition signal \cite{Rischke95} in the 
excitation function of directed flow.

2. The {\em elliptic flow} is strongest near midrapidity \cite{NA49flow}.
Its driving force is the azimuthal anisotropy of the transverse pressure
gradient, caused by the geometric deformation of the reaction region
in the transverse plane. As pointed out in \cite{S99,KSH99}, elliptic
flow acts against its own cause by eliminating the geometric deformation
which generates it, thereby shutting itself off after some time. This time 
is, at least at high energies, longer than the nuclear transition time.
Elliptic flow is thus generated {\em later} than directed flow, and 
hydrodynamic concepts for its description may have a larger chance of 
being valid, even if the spatial deformation which causes elliptic 
flow exists only for a fraction of the total fireball lifetime. An 
important focus of this work will be a quantitative determination 
of the time scale over which elliptic flow is generated, as a 
function of the collision energy. We will see that this time scale 
grows with the overall size of the (initially deformed) collision region 
\cite{O92,O98}. Studying central collisions between large deformed nuclei 
like $^{238}$U \cite{Sh00,Li00} therefore improves the chances that 
thermalization happens sufficiently early for a hydrodynamic description 
of elliptic flow evolution to be valid. Such collisions are the 
preferred proving ground for hydrodynamic predictions for the 
excitation function of elliptic flow.

3. {\em Radial flow} is generated by the pressure gradient between the
interior of the collision fireball and the external vacuum; this force
persists throughout the fireball expansion until freeze-out. Of all
three transverse flow patterns it thus has the strongest weight at
late times. Comparing the excitation functions of elliptic and radial
flow with their intrinsically different weights for the EOS at early 
and late times (i.e. at high and low energy density) may help with 
the identification of phase transition signatures and their 
discrimination against possible non-equilibrium effects from incomplete 
local thermalization. Of course, the final proof for the phase transition
to quark matter will require an additional correlation of the here
predicted structures in the anisotropic flow pattern with other
``quark-gluon plasma signatures'' (see \cite{BGHG99,HJ00}).

As already indicated we here study the evolution of transverse flow
in a macroscopic hydrodynamic framework (to be contrasted with
microscopic kinetic approaches \cite{S97,S99,S97a}). This approach, 
which is based on the assumption of rapid local thermalization,
allows the most direct connection of observables to the EOS of the hot 
matter in the collision fireball, including possible
phase transitions. Its validity can be tested both experimentally and
by comparison with kinetic approaches. We will not do so here (see, 
for example, Refs. \cite{HL99,VP00}) but rather concentrate on 
qualitative predictions resulting from the hydrodynamic approach. 

Hydrodynamics cannot describe the earliest collision stage of nuclear 
energy loss and entropy production by thermalization of the energy 
deposited in the reaction zone by the stopping process; this must be 
replaced by appropriate initial conditions for the hydrodynamic expansion.
The evolution of azimuthally asymmetric reaction zones requires a 
(3+1)-dimensional hydrodynamic approach. This is very time consuming 
and makes a tuning of initial conditions to data difficult 
\cite{Hirano,Netal99}. However, 
near midrapidity and especially for high collision energies the 
longitudinal expansion dynamics is expected to be given by the 
Bjorken scaling solution \cite{Bj83} which can be implemented
analytically. The remaining hydrodynamic equations for the transverse
dynamics live in 2 space and 1 time dimension and are much easier to 
solve \cite{O92,KSH99,TS99}. The hydrodynamic evolution is terminated 
by a freeze-out criterium (in our case a fixed decoupling energy 
density). At this point the energy and baryon densities are converted 
into temperature and chemical potentials for baryon number and 
strangeness, using the EOS, and the particle spectra are calculated 
using the Cooper-Frye prescription \cite{CF74}. With these spectra 
and the hydrodynamic flow pattern on the freeze-out surface the 
average radial flow velocity $\lda v_\perp \rda$ and the elliptic 
flow coefficient $v_2$ are evaluated.

The present paper gives technical details for our previous two short
reports in \cite{KSH99} and significantly extends the results presented 
there. The excitation function for $v_2$ is complemented by a similar 
one for the average radial flow and calculated up to very much higher 
energies. We also compute the impact parameter dependence at fixed 
beam energy of the elliptic flow scaled by the initial spatial 
anisotropy. The time evolutions of radial and elliptic flow and
their dependence on the collision energy are discussed in detail, in 
order to establish to which extent elliptic flow is really a signature
for {\em early pressure} in the system \cite{S97,S99,KSH99}.
Finally, we advertize central U+U collisions in the side-on-side 
configuration as an optimum system for studying the hydrodynamic 
evolution of elliptic flow and the quark-hadron phase transition
signature in its beam energy dependence \cite{KSH99}. We give predictions
for the time evolution of radial and elliptic flow, for their excitation
function and for the $p_{_{\rm T}}$-dependence of the elliptic flow 
coefficient at SPS energies for this particular collision system.

\se{The hydrodynamic model}
\label{sec2}

The equations of relativistic ideal hydrodynamics follow from the local
conservation laws for energy, momentum, and other conserved currents
(e.g. baryon number),
 \beq{econs}
   {\partial_{\mu} T^{\mu \nu}}(x)=0
   \mbox{\hspace{.6cm} and \hspace{.6cm}}
   \partial_\mu j^\mu(x) =0\,, 
 \eeq
by inserting the ideal fluid decompositions
 \bea{decomp1}
   &&T^{\mu\nu}(x)=\Bigl(e(x)+p(x)\Bigr) u^\mu(x) u^\nu(x) 
     - g^{\mu\nu} p(x)\,,
 \\
 \label{decomp2}
   &&j^\mu(x) = n(x)\, u^\mu(x)\,.
 \eea
$e(x)$ is the energy density, $p(x)$ the pressure, and $n(x)$ the 
conserved number density at point $x^\mu{=}(t,x,y,z)$; 
$u^\mu(x){\,=\,}\gamma (1,v_x,v_y,v_z)$ with 
$\gamma{\,=\,}1/{\textstyle\sqrt{1{-}v_x^2{-}v_y^2{-}v_z^2}}$
is the local four velocity of the fluid. Ideal hydrodynamics assumes
that local thermalization by the strong interactions among the matter 
constituents happens fast on the scale defined by the space-time 
gradients of these quantities and therefore neglects such 
gradient terms \cite{Cs94}.

We always use $x$ for the transverse coordinate inside the reaction 
plane, with positive values in the direction of the impact parameter
$\bbox{b}$, and $y$ for the transverse coordinate perpendicular to
$\bbox{b}$. (In momentum space $y$ denotes the rapidity; which meaning
is implied should follow from the context.) $z$ points in beam direction.

\suse{The equation of state}
\label{sec2a}

The set (\ref{econs}) contains 5 equations for 6 unknown fields 
$e, n, p, v_x, v_y, v_z$. To close the system one needs an equation
of state (EOS) which relates pressure, energy and baryon density.
The EOS for strongly interacting matter involves a phase transition 
from a hadron resonance gas (HG) phase to a color-deconfined 
quark-gluon plasma (QGP) phase. Like many others before (see, e.g., 
\cite{LRH86,Setal97}) we accomplish this by separately constructing 
an EOS for a resonance gas (EOS~H) and for the QGP phase (EOS~I)
%
 \begin{figure}[h,t,b,p]
 \epsfig{file=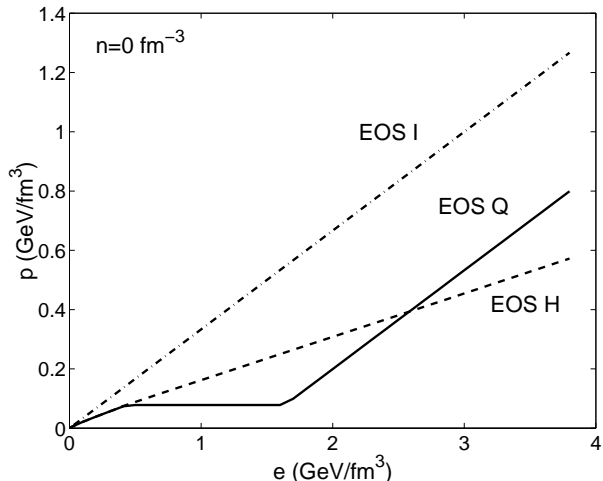,width=8cm}
 \caption{The three equations of state discussed in the text, at 
          vanishing net baryon density.}
 \label{F1}
 \end{figure}
%
\noindent
and matching the two via the Maxwell construction, invoking a bag constant
$B$ to describe the different vacuum energy in the two phases. EOS~H is 
constructed from the contributions of all known hadron resonances of 
masses up to 2 GeV; their repulsive short-range interactions are 
parametrized via a mean-field potential ${\mathcal V}(n)=\half K n^2$ 
with $K=0.45$ GeV\,fm$^3$ \cite{Setal97}. The QGP is described as an 
ideal gas of massless quarks and gluons (EOS~I) inside a large bag with 
bag constant $B$. The latter is tuned to the desired phase transition 
temperature: $B^{1/4}=230$\,MeV gives $T_{\rm c}(n=0) = 164$\,MeV at 
vanishing net baryon density. EOS~I is given by the simple equation 
$p(e,n) = {1\over 3} e$ or $\P p/\P e = {1\over 3}$, independent of 
$n$.

\begin{figure}[h,t,b,p]
 \epsfig{file=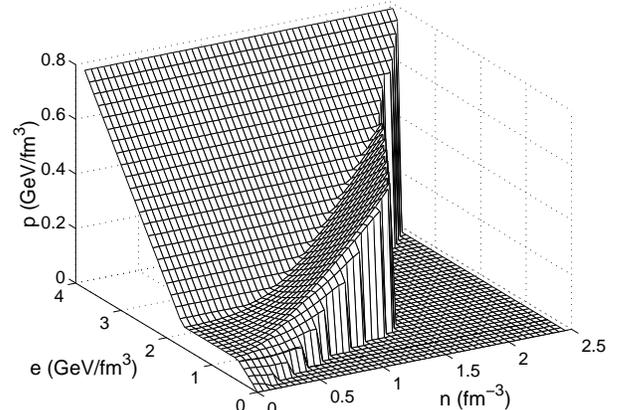,width=8cm}
 \caption{The equation of state EOS~Q with a quark-hadron phase 
    transition. The pressure is shown as a function of energy and
    net baryon density, $e$ and $n$. For each value of $n$ there exists 
    a minimum energy density $e_{\rm min}(n)$ with corresponding pressure
    $p_{\rm min}(e_{\rm min},n)$; below $e_{\rm min}$ the pressure is set 
    to zero by hand.}
 \label{F2}
 \end{figure}

In order to investigate the influence of the phase transition on 
the anisotropic transverse flow pattern, we stu\-died separately the 
equations of state EOS~H and EOS~I as well as the combined equation 
of state EOS~Q which includes the phase transition between the first 
two as obtained from the Maxwell construction. Comparisons to data 
are only performed for EOS~Q. Figure \ref{F1} shows all three 
equations of state for vanishing net baryon density $n=0$ while 
Fig.~\ref{F2} gives for EOS~Q the pressure as a function of both 
$e$ and $n$.

\suse{Reduction to 2+1 dimensions}
\label{sec2b}

At high collision energies, relativistic kinematics and its influence 
on the particle production process implies longitudinal boost-invariance 
of the collision fireball near midrapidity \cite{Bj83}. (Of course, 
near the target and projectile rapidities longitudinal boost-invariance 
is broken by the finite amount of total available energy.) As a result,
the longitudinal velocity field scales as $v_z=z/t$, and it is convenient 
to use a coordinate system spanned by longitudinal proper time 
$\tau=t\sqrt{1-v_z^2}$ and space-time rapidity $\eta = \half
\ln[(t{+}z)/(t{-}z)]$ instead of $t$ and $z$ (see Appendix~\ref{appa}).
Longitudinal boost-invariance is then equivalent with 
$\eta$-independence.

Assuming the validity of this scaling ansatz near midrapidity, the
longitudinal expansion of the fireball can be dealt with analytically,
thereby reducing the numerical problem to the two transverse dimensions
and time \cite{O92}. This greatly reduces the numerical effort. However,
by doing so one gives up the possibility to study the rapidity dependence
of the (anisotropic) transverse flow pattern \cite{Hirano,Netal99}
as well as other interesting effects which occur at AGS and SPS 
energies, like the tilt of the longitudinal axis of the collision 
fireball away from the beam direction \cite{Netal99,Betal00,LHW00}. For 
such studies a complete solution of the (3+1)-dimensional hydrodynamics 
\cite{Hirano,Netal99,Betal00,RBM95} is required. We will here concentrate
entirely on the midrapidity region where the (2+1)-dimensional approach 
with exact longitudinal boost-invariance is expected to yield reasonable
results even at SPS energies. At higher energies the model should become
better and better.

The implementation of longitudinal boost-invariance and transformation
from $(t,z)$ to $(\tau,\eta)$ is described in Appendix~\ref{appa}. The
rewritten hydrodynamic equations read
 \bea{DGL}
   \begin{array}{c@{\,+\;}c@{\,+\;}c@{\,=\;}l}
   \partial_\tau\tilde{T}^{\tau\tau}                           &   
   \partial_x\left(\tilde{v}_{x} \tilde{T}^{\tau\tau}\right)   &   
   \partial_y\left(\tilde{v}_{y} \tilde{T}^{\tau\tau}\right)   &    
   -p\,,
 \\
   \partial_\tau\tilde{T}^{\tau x}                             &
   \partial_x\left(\bar{v}_x \tilde{T}^{\tau x}\right)         &
   \partial_y\left(\bar{v}_y \tilde{T}^{\tau x}\right)         & 
  -\partial_x\tilde{p}\,,
 \\
   \partial_\tau\tilde{T}^{\tau y}                             &
   \partial_x\left(\bar{v}_x \tilde{T}^{\tau y}\right)         &
   \partial_y\left(\bar{v}_y \tilde{T}^{\tau y}\right)         &
  -\partial_y\tilde{p}\,,
 \\
   \partial_\tau\tilde{\jmath}^\tau                            &
   \partial_x(\bar{v}_x \tilde{\jmath}^\tau)                   &
   \partial_y(\bar{v}_y \tilde{\jmath}^\tau)                   &
   0\,,
 \end{array}
 \eea
where 
 \bea{defns}
   &&\tilde T^{\mu \nu} = \tau T^{\mu \nu}\,,\quad
     \tilde p = \tau\, p\,,
 \\
   &&\bar v_i = v_i \cosh\eta\,, \quad
     \tilde v_i = {T^{\tau i}\over T^{\tau \tau}}
                = {(e+p)\bar\gamma^2\bar{v}_i \over
                   (e+p)\bar\gamma^2{-}p}
     \,,\quad(i=x,y).
 \nonumber
 \eea   
We call $\bar v_i$ the transport velocities and $\tilde v_i$ the energy
flow velocities in the transverse directions. Since we work at midrapidity, 
$\eta=0$, the transverse transport velocities agree with the corresponding
fluid velocities in the c.m. frame.

In hydrodynamic problems phase transitions generically lead to the 
formation of shock waves which complicate the numerical solution.
To integrate the differential equations (\ref{DGL}) we use the 
``Sharp and Smooth Transport Algorithm'' (SHASTA \cite{BB73}) 
which was shown to perform excellently even under difficult 
conditions \cite{RBM95}.

\suse{Initialization of the fields}
\label{sec2c}

In this subsection we discuss the initial conditions for the solution
of Eqs.~(\ref{DGL}). Strong interactions between the partons of the 
colliding nuclei lead to the deposition of a large fraction of the 
beam energy and the creation of many secondary particles in the 
reaction zone. The newly produced partons interact strongly with each 
other and, after only a few scatterings during a time interval 
$\tau_0 = {\cal O}$(1\,fm/$c$), the system is expected to reach a 
state of approximate local thermal equilibrium. Following 
\cite{O92,BO90} (to which we refer for details) we take the energy 
deposition in the transverse plane to be proportional (by a factor $K$)
to the number of collisions producing wounded nucleons:
 \bea{init}
   \FL
   && e(x,y;\tau_0) =
 \\
   && K \left\{ T_A\bigl(x{+}{\textstyle{b\over 2}},y\bigr)
      \Bigl[1-\Bigl(1-{\sigma T_B\bigl(x{-}{\textstyle{b\over 2}},y\bigr)
                       \over B}\Bigr)^B \Bigr] \right. 
 \nonumber\\
   &&\ \ + \left. T_B\bigl(x{-}{\textstyle{b\over 2}},y\bigr)
     \Bigl[1-\Bigl(1-{\sigma T_A\bigl(x{+}{\textstyle{b\over 2}},y)
                      \over A}\Bigr)^A \Bigr]\right\}\,.
 \nonumber
 \eea
Here $T_A$ is the nuclear thickness function of the incoming 
nucleus $A$,
 \beq{TA}
   T_A(x,y)=\int_{-\infty}^{+\infty} dz\, \rho_A(x,y,z) \,,
 \eeq
where the nuclear density $\rho_A$ is given by a Woods-Saxon profile,
 \beq{rhoA}
   \rho_A(\bbox{r}) = \frac{\rho_0}{1+\exp[(r-R_0)/\xi]}\,,
 \eeq
and similarly for nucleus $B$. 

We further assume that the initial transverse density profile of net 
baryon number is proportional to the initial transverse energy 
density profile:
 \beq{n}
   n(x,y;\tau_0)=L\,e(x,y;\tau_0)\,.
 \eeq
For Pb-Pb collisions we use in (\ref{rhoA}) a nuclear radius 
$R_0=6.5$\,fm and a surface thickness $\xi=0.54$\,fm \cite{BM69}. 
For U-U collisions we take $R_0=6.8$\,fm, with a deformation 
$\delta=0.27$ (\cite{BM69}, Vol.~2, p.~133). This leads to a 
ratio $R_l/R_s=1.29$ between the long and short axes of this 
nucleus; in absolute terms $R_l=8.0$\,fm and $R_s=6.2$\,fm \cite{fn1}. 
For the ground-state nuclear density we take $\rho_0=0.17$\,fm$^{-3}$ 
\cite{BM69}. 

\begin{figure}[h,t,b,p]
\hspace*{-0.4cm}\epsfig{file=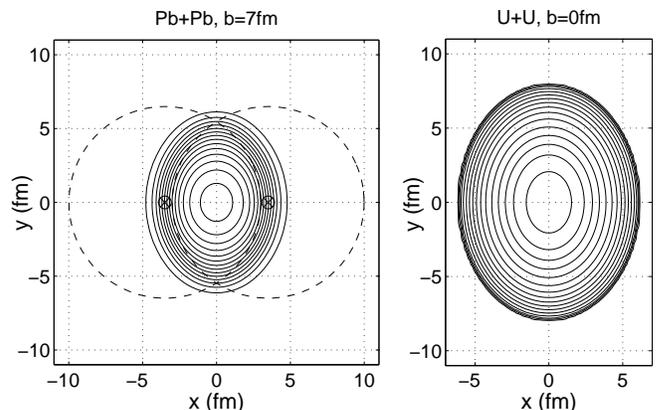,width=\linewidth}
\caption{Left: initial transverse energy density distribution for a 
   typical 158\,$A$\,GeV/$c$ Pb+Pb collision at impact parameter 
   $b{\,=\,}7$\,fm. Indicated are contours of constant ener\-gy density 
   between $e{\,=\,}7.0$\,GeV/fm$^3$ (innermost contour) and 
   $e{\,=\,}0.5$\,GeV/fm$^3$ (outermost contour) in steps of 
   $\Delta e$\,=\,0.5 GeV/fm$^3$. The dashed lines represent the 
   colliding nuclei before impact. Right: The same for a central 
   155\,$A$\,GeV/$c$ side-on-side U+U collision -- the innermost 
   (outermost) contour corresponds to $e{\,=\,}8.0$\,GeV/fm$^3$ 
   (0.5\,GeV/fm$^3$). 
\label{F3}}
\end{figure}

Three parameters thus describe the initial conditions:
\bit
\item{the maximum energy density $e_0$ in a central collision 
      ($b=0$) -- this fixes the parameter $K$ in (\ref{init}) at 
      the given beam energy;}
\item{the ratio $L$ in (\ref{n}) between energy and baryon density;}
\item{the equilibration time $\tau_0$.}
\eit
In Sec.~\ref{sec2e} we adjust the parameters by tuning the output of 
our calculations with EOS~Q for central ($b=0$) Pb+Pb collisions to 
experimental data (transverse mass spectra of negative hadrons and 
net protons at midrapidity \cite{NA49spec}) at 158 $A$\,GeV/$c$ beam 
momentum. We use the same parameters $K,L$ and $\tau_0$ for U+U 
collisions at 155 $A$\,GeV/$c$.

In Fig.~\ref{F3} we illustrate the initial conditions resulting from 
this tuning procedure. It shows contour plots of the energy density 
in the transverse plane at $z=0$ for Pb+Pb collisions with $b=7$\,fm 
and central U+U collisions in the side-on-side configuration at the 
highest SPS beam momentum of $400\times(Z/A)$ GeV/$c$. Note that at 
fixed collision energy the central energy density for $b{\,=\,}0$ 
side-on-side U+U collisions is 8\% lower than for $b{\,=\,}0$ Pb+Pb 
collisions, but about 14\% higher than in Pb+Pb collisions at 
$b{\,=\,}7$\,fm which correspond to about the same initial spatial 
deformation. At similar deformation, the initial volume of the 
elliptic fireball formed in central side-on-side U+U collisions is 
almost twice that of the corresponding semi-central Pb+Pb collisions.

\suse{Freeze-out and particle spectra}
\label{sec2d}

As the matter expands and cools, the mean free path of the matter
constituents grows, and the hydrodynamical description eventually
breaks down. The system reaches the point of ``kinetic freeze-out''
after which the momentum spectra are no longer significantly affected 
by scattering among the particles. One should stop the hydrodynamic 
solution when the average time between scatterings $\tau_{\rm scatt}=
 1/\la v\sigma \ra n$ becomes comparable to the expansion time scale 
$\tau_{\rm exp} = 1/\partial\cdot u$ (inverse ``Hubble constant'') 
\cite{BGZ78,HLR87,LHS89}. [It was shown in \cite{SH92} that in 
relativistic heavy ion collisions freeze-out happens {\em dynamically} 
rather than geometrically, i.e. it is driven by the expansion of the 
fireball and not by its finite size.] Numerical calculations 
\cite{SH92,MH97} have shown that, since the particle density in 
the denominator of $\tau_{\rm scatt}$ is a very steep function of 
$T$, this leads to freeze-out at nearly constant temperature. For 
low net baryon freeze-out densities, as they arise in heavy ion 
collisions at and above SPS energies near midrapidity, this corresponds 
to almost constant energy density. We here therefore impose freeze-out
at a constant energy density $e_{\rm dec}$ which is the most easily 
implemented condition in hydrodynamics. The value of $e_{\rm dec}$ (or,
nearly equivalently, $T_{\rm dec}$) is another model parameter to be 
tuned to the data. 

After the freeze-out hypersurface $\Sigma$ of constant energy density
$e_{\rm dec}$ has been determined, the $T_{\rm dec}(x)$, chemical 
potentials $\mu_i(x)$ and flow velocity field $u_\mu(x)$ are evaluated 
on this surface. To this end a tabulated version of EOS~H is used
for interpolation which (in addition to the pressure $p$) gives the 
intensive thermodynamical variables as functions of $e$ and $n$.
Each cell $x$ on this freeze-out hypersurface contributes particles
of species $i$ (where $i$ runs over all resonances included in EOS~H)
with a local equilibrium distribution
 \beq{distr}
   f_i(x,p) = {g_i\over (2\pi)^3} \, 
   {1\over e^{[p\cdot u(x) - \mu_i(x)]/T_{\rm dec}(x)}\pm 1} \,.
 \eeq
$g_i$ is the spin-isospin degeneracy factor for particle species $i$.
The complete momentum spectrum is obtained by summing the corresponding
particle flux currents across the 3-dimensional freeze-out hypersurface 
$\Sigma$ in space-time over all cells in $\Sigma$ (Cooper-Frye 
prescription \cite{CF74}):
 \beq{CF} 
   E {dN_i\over d^3p} = 
   {dN_i \over dy\, p_{_{\rm T}} dp_{_{\rm T}} \, d\phi} =
   \int_\Sigma p\cdot d^3\sigma(x)\, f_i(x,p)\,. 
 \eeq
This prescription is strictly correct only for freeze-out surfaces
whose normal vector $d^3\sigma(x)$ is everywhere timelike because 
otherwise some particles flow back into the 4-volume inside 
$\Sigma$. A discussion of this issue which still awaits 
a fully consistent solution can be found in \cite{BMGR96,Bu96}.

In the present paper we concentrate on flow patterns reflected in pion 
spectra. (Flow anisotropies for pions and protons at SPS energies were 
compared in \cite{KSH99}.) A significant fraction of the measured pions 
arises from the decays of unstable resonances after freeze-out. These
decays usually happen isotropically in the rest frame of the resonance
and tend to smear out flow anisotropies, thereby reducing the anisotropic 
flow signals \cite{KSH99,Hirano}. The fraction of pions from resonance 
decays depends strongly on the freeze-out temperature: their diluting
effect on the elliptic flow $v_2$, for example, is much stronger at 
$T_{\rm dec}{\,=\,}140$\,MeV \cite{Hirano} than at 
$T_{\rm dec}{\,=\,}120$\,MeV \cite{KSH99}. All our calculations fully
account for resonance decay contributions including the complete
relativistic decay kinematics \cite{SKH91}. 

\suse{Tuning the model}
\label{sec2e}
 
Since the hydrodynamic approach cannot describe the initial thermalization
stage directly after nuclear impact, the initial conditions for the
hydrodynamic expansion stage cannot be predicted but must be obtained
by fitting experimental data. However, once the initial conditions 
(in our case the parameters $K,L$, and $\tau_0$) have been fixed in 
central collisions, the Glauber model (\ref{init}) uniquely predicts 
their dependence on the impact parameter. The validity of the 
hydrodynamic model can thus be tested by checking the impact parameter
dependence of its predictions. In \cite{KSH99,Kolb} we showed that, 
after being tuned to central Pb+Pb collisions at 158 $A$\,GeV/$c$, the 
model successfully reproduces the measured pion and proton spectra near 
midrapidity up to impact parameters of 8-10\,fm. This was better than 
expected.

\begin{figure}[htbp]
\epsfig{file=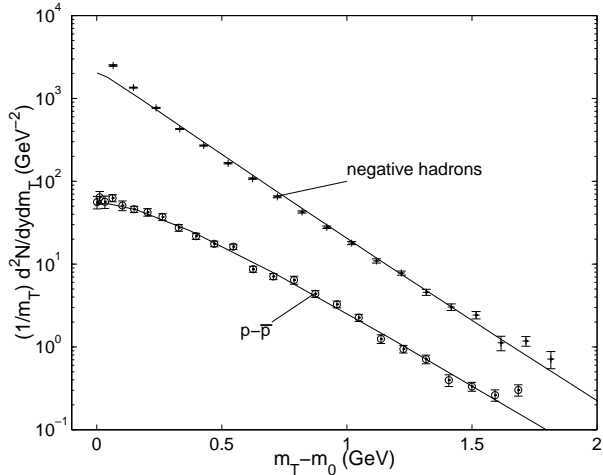,width=8cm}
\caption{Particle spectra from central Pb+Pb collisions at 
  158\,$A$\,GeV/$c$ at midrapidity \protect\cite{NA49spec} together 
  with the hydrodynamical model predictions after tuning of the model 
  parameters (solid lines).
\label{F4}}
\end{figure}

We here provide some details of the tuning procedure which were not
previously reported in \cite{KSH99} due to space limitations. In 
particular we show in Fig.~\ref{F4} our fit to the midrapidity 
$m_{_{\rm T}}$-spectra of negative hadrons ($h^-$) and net protons 
measured by the NA49 collaboration \cite{NA49spec}. The theoretical 
spectra are absolutely normalized. The corresponding fit parameters for 
the initial state are $e_0{\,=\,}9.0$ GeV/fm$^3$ for the initial energy 
density in the center of the fireball (corresponding to $K=2.04$\,GeV/fm 
in (\ref{init}) and to an initial central temperature $T_0{\,=\,}258$ 
MeV \cite{fn1a}), $n_0{\,=\,}1.1$ fm$^{-3}$ for the initial baryon density 
in the fireball center (corresponding to $L=0.122$\,GeV$^{-1}$ in (\ref{n})), 
and a starting time $\tau_0=0.8$\,fm/$c$ for the hydrodynamic expansion
(corresponding to $T_0\cdot\tau_0/\hbar{\,=\,}1.05$). $\tau_0$ controls 
the dilution of the matter via boost-invariant longitudinal expansion 
and thus the length of time available for the buildup of transverse 
flow before freeze-out; the latter affects the slope of the 
$m_{_{\rm T}}$-spectra. The total time until freeze-out and the amount 
of transverse flow generated can also be changed by varying the initial 
energy density, but this also affects the normalization of the
midrapidity spectra. $e_0$ and $\tau_0$ result from a suitable 
balance between these two effects. $n_0$ is then essentially fixed 
by the measured ratio between the proton and $h^-$ spectra.

The different shapes of the proton and $h^-$ spectra provide a handle
to separate collective transverse flow ($\lda v_\perp \rda$) from 
thermal motion ($T_{\rm dec}$) at freeze-out. However, it is known that
a thermal model analysis of particle spectra in general results in 
strong correlations between these two parameters \cite{LHS89,SH92}.
Our best fit gives $T_{\rm dec}{\,\approx\,}120$\,MeV (corresponding 
to $e_{\rm dec}{\,=\,}0.06$ GeV/fm$^3$) and 
$\lda v_\perp\rda{\,=\,}0.45\,c$, albeit with a significant 
uncertainty (somewhat lower $T_{\rm dec}$ with higher $\lda v_\perp \rda$ 
and vice versa cannot be excluded). This is in good agreement with 
other analyses of particle spectra \cite{Ketal97} and hydrodynamic
simulations \cite{HS98}; a combined analysis of spectral slopes and 
two-particle Bose-Einstein correlations \cite{NA49HBT,TWH99}
tends to give somewhat larger transverse flow velocities coupled to 
lower freeze-out temperatures, but still inside the region of 
uncertainty from the analysis of the single-particle spectra.

This set of fit parameters, adjusted to SPS data, is our starting 
point for extrapolations towards non-central collisions and into 
different collision energy regimes. When studying the impact parameter 
dependence at fixed collision energy we leave all parameters unchanged. 
This may be unrealistic for very peripheral collisions where the 
midrapidity fireball is smaller and geometric freeze-out can cut the 
expansion short, leading to higher decoupling temperatures. For the
spectral slopes this is a second order effect since earlier freeze-out
at higher $T_{\rm dec}$ is partially compensated for by a smaller 
transverse flow velocity $\lda v_\perp \rda$. As we will see below
(see Fig.~\ref{F7}), the elliptic flow anisotropy $v_2$ builds up early 
in the collision and, even at SPS energies, has almost reached its 
final value already several fm/$c$ before decoupling; a possible 
earlier decoupling in very peripheral collisions thus will not 
strongly affect $v_2$ either. We thus feel justified in leaving the 
model parameters (in particular the decoupling temperature) unchanged 
when studying the impact parameter dependence.

When investigating the excitation function of radial and elliptic
flow we change $K$ and $\tau_0$. This is rationalized as follows:
At higher energies we expect higher particle production per 
wounded nucleon; we cannot predict the beam energy dependence of 
secondary particle production, but we can parametrize it by changing 
$K$ and plotting our results as a function of the finally observed 
multiplicity density $dN/dy$. The beam energy dependence of $dN/dy$
will eventually be provided by experiment, then allowing to present
our results directly against $\sqrt{s}$. -- Higher initial particle
production leads to higher particle and energy densities and thereby
to accelerated thermalization. From relativistic kinematics and the 
uncertainty relation it follows that the production time of a secondary 
particle is inversely related to its energy \cite{KLS92}; by dimensional 
analysis this suggests that the thermalization time $\tau_0$ scales in 
inverse proportion to the initial temperature $T_0$: 
$T_0 \tau_0$\,=\,const. or, equivalently, $\tau_0\,K^{1/4}$\,=\,const. 
This is what we use in the present paper; in \cite{KSH99} we instead 
left $\tau_0$ constant. Within the range of collision energies studied 
in \cite{KSH99} the difference is negligible, but for the higher 
energies investigated here a reduction of $\tau_0\sim 1/T_0$ causes
a significant shrinkage of the horizontal axis on the excitation 
function in Fig.~\ref{F14} below.

For energies above the SPS we leave the initial baryon density
$n(x,y;\tau_0)$ unchanged. As a result, the ratio $L$ of baryon to
energy density drops, qualitatively consistent with the expectation of
decreasing baryon stopping. Since already at the SPS the influence of 
the baryons on the EOS is minor, it doesn't really matter in which way
$L$ approaches zero as the collision energy goes to infinity. Note
that we don't predict the normalization of the baryon spectra at other
than SPS energies. Below SPS energies we leave $L$ constant, lacking
motivation for a different choice. Once a better understanding of the
beam energy dependence of the initial conditions becomes available,
this can be easily improved.  
  
\se{Transverse flow phenomenology}
\label{sec3}

In this section we study generally the space-time evolution of the 
transverse flow pattern and how it is influenced by a phase transition
in the EOS. Since the finally observed particle spectra and their azimuthal
anisotropies reflect the full space-time history of the fireball expansion,
their proper interpretation requires an accurate understanding of the 
transverse fireball evolution. In \cite{KSH99} we showed that the 
softening of the EOS in the phase transition region leads at collision 
energies above the SPS to a reduction of the elliptic flow coefficient 
$v_2$ below the value expected from a hadron resonance gas. At even higher
energies, however, one expects to enter a regime where the initial 
energy density is so far above the phase transition that nearly all of
the expansion history happens inside the QGP phase. Since far above 
$T_{\rm c}$ the EOS of a QGP ($p{\,=\,}{1\over 3}e-B$) is much harder
than EOS~H (which in the region relevant for us can be parametrized
by $p{\,\approx\,}0.15\,e$), $v_2$ should eventually rise again and 
approach the value characteristic of EOS~I which is 30-40\% higher. 
In order to see whether this is true we have now studied collisions at
very much higher energies, even far beyond the LHC.
 
\suse{Semiperipheral Pb+Pb collisions}
\label{sec3a}

\begin{figure}[htbp]
\bce
\epsfig{file=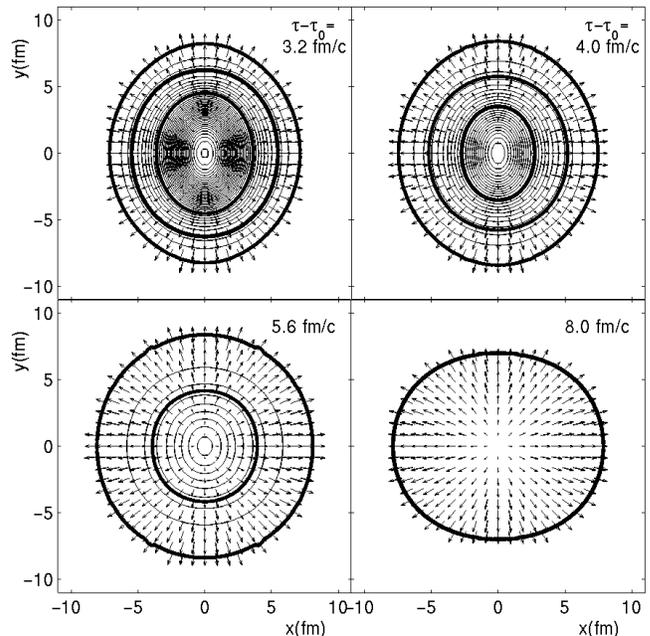,width=8.5cm}
\caption{Time evolution for EOS~I of the transverse energy density 
   profile (indicated by constant energy density contours spaced by 
   $\Delta e{\,=\,}150$\,MeV/fm$^3$) and of the flow velocity 
   field (indicated by arrows) for Pb+Pb collisions at impact 
   parameter $b{\,=\,}7.0$\,fm. The four panels show snapshots 
   at times $\tau{-}\tau_0{\,=\,}3.2$, 4.0, 5.6, and 8.0\,fm/$c$. At
   these times the maximal energy densities in the center are 5.63,
   3.62, 1.31 and 0.21 GeV/fm$^3$, respectively. For further details
   see text.  
\label{F5}}
\ece
\end{figure}

In this subsection we investigate Pb+Pb collisions at an impact
parameter of 7\,fm (left panel in Fig.~\ref{F3}). We begin by showing
the evolution of the energy distribution and flow field in the
transverse $(x,y)$ plane for the cases with and without a phase
transition. We do so for an initial central energy density in
$b{\,=\,}0$ Pb+Pb collisions of $e_0{\,=\,}175$\,GeV/fm$^3$
($T_0{\,=\,}510$\,MeV) at $\tau_0{\,=\,}0.38$\,fm/$c$. The resulting
total pion multiplicity density with EOS~Q of 
${dN_\pi\over dy}\big\vert_{y=0}{\,=\,}1070$ at $b{\,=\,}7$\,fm is at
the upper end of the range of predictions for RHIC energies \cite{RHIC}. 
This study was motivated by the work of Teaney and Shuryak who
predicted under similar conditions an interesting phenomenon which
they called ``nutcracker flow'' \cite{TS99} and which shows up only in
the presence of a phase transition. In Fig.~\ref{F5} we show the
evolution for EOS~I, i.e. a hard EOS without phase transition. One 
sees smooth expansion and a continuous transition from an initial
state of positive elliptic deformation (longer axis perpendicular to
the collision plane) to one with negative deformation, caused by the 
developing in-plane elliptic flow. The thicker contours correspond 
(from the inside outward) to $e{\,=\,}1.6$, 0.45, and 0.06\,GeV/fm$^3$; 
for the more realistic equation of state EOS~Q the first two values 
limit the mixed phase while the latter indicates freeze-out.

\begin{figure}[htbp]
\bce
\epsfig{file=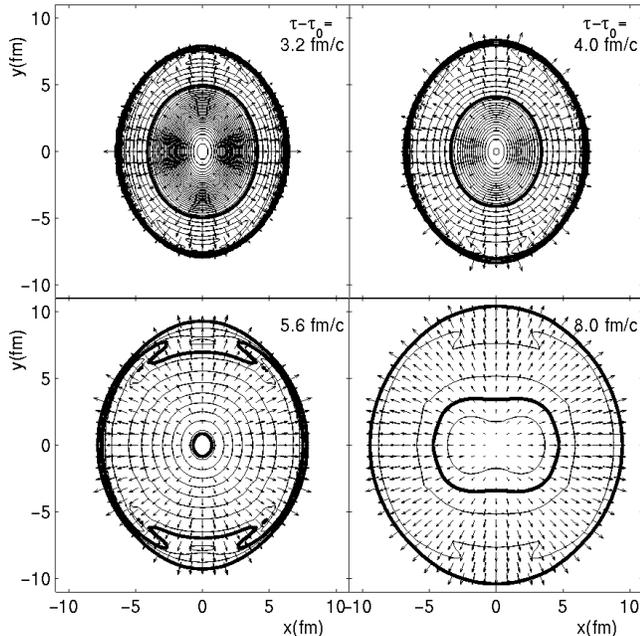,width=8.5cm}
\caption{Same as Fig.~\ref{F5}, but for EOS~Q which features a phase
   transition. The spacing between energy density contours is again 
   150 MeV/fm$^3$, and the snapshots are taken at the same times. The 
   corresponding maximum energy densities are 5.97, 3.97, 1.67, and
   0.55 GeV/fm$^3$, respectively. See text for discussion.
\label{F6}}
\ece
\end{figure}

\begin{figure}[htbp]
\bce
\hspace*{0.3cm}\epsfig{file=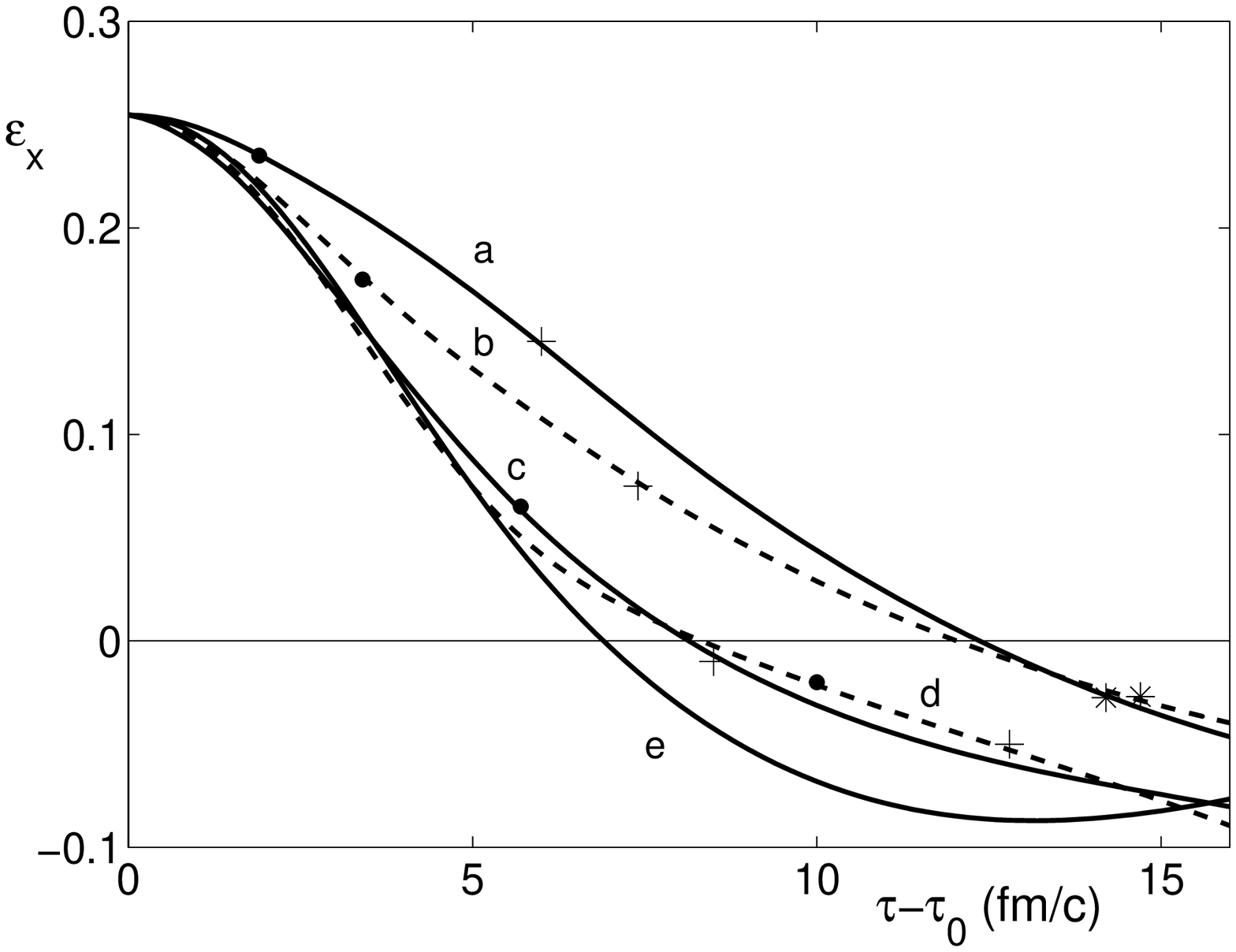,height=4.9cm,width=7.1cm}\\
\hspace*{0.1cm}\epsfig{file=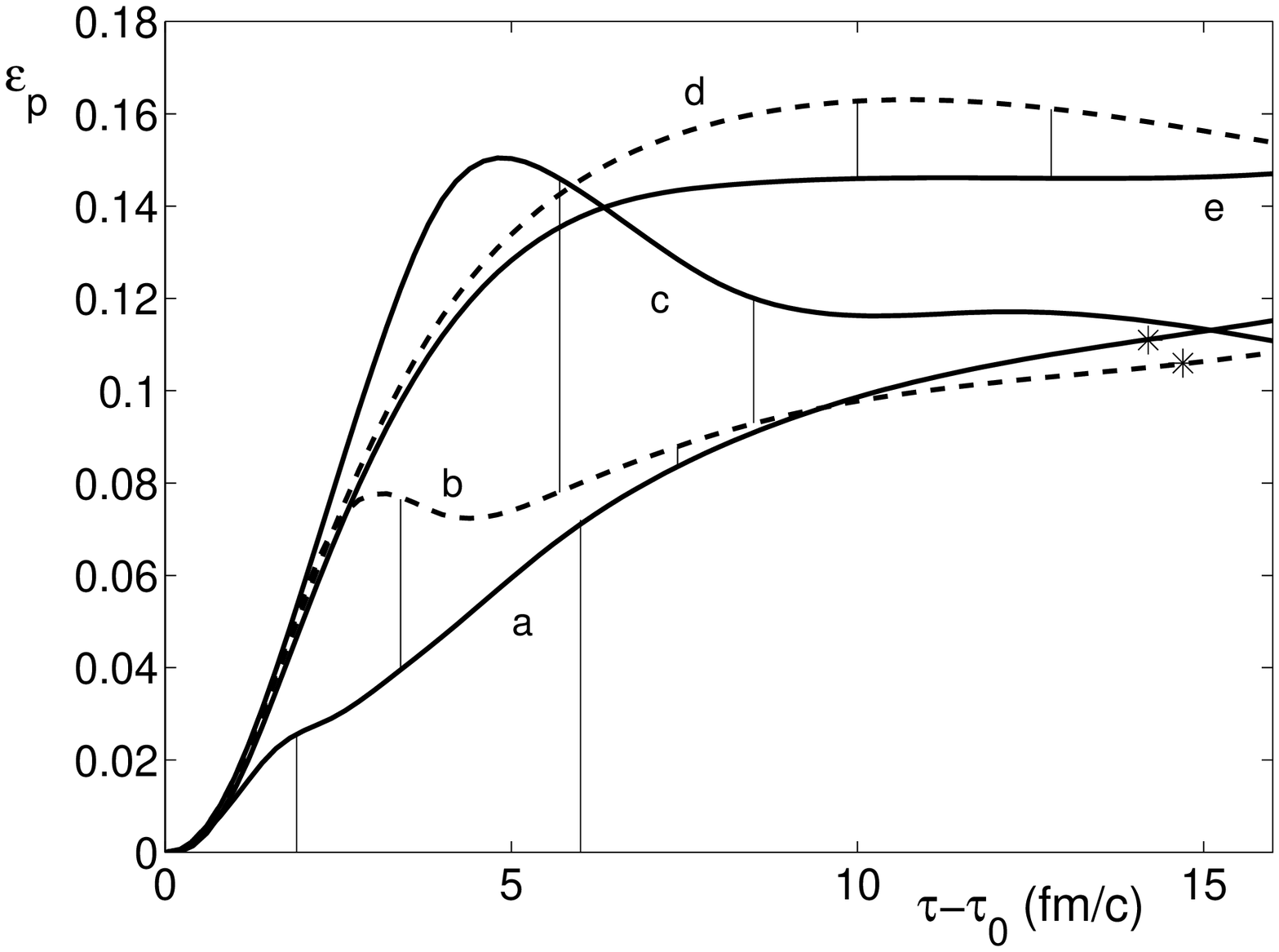,height=4.9cm,width=7.34cm}\\
\epsfig{file=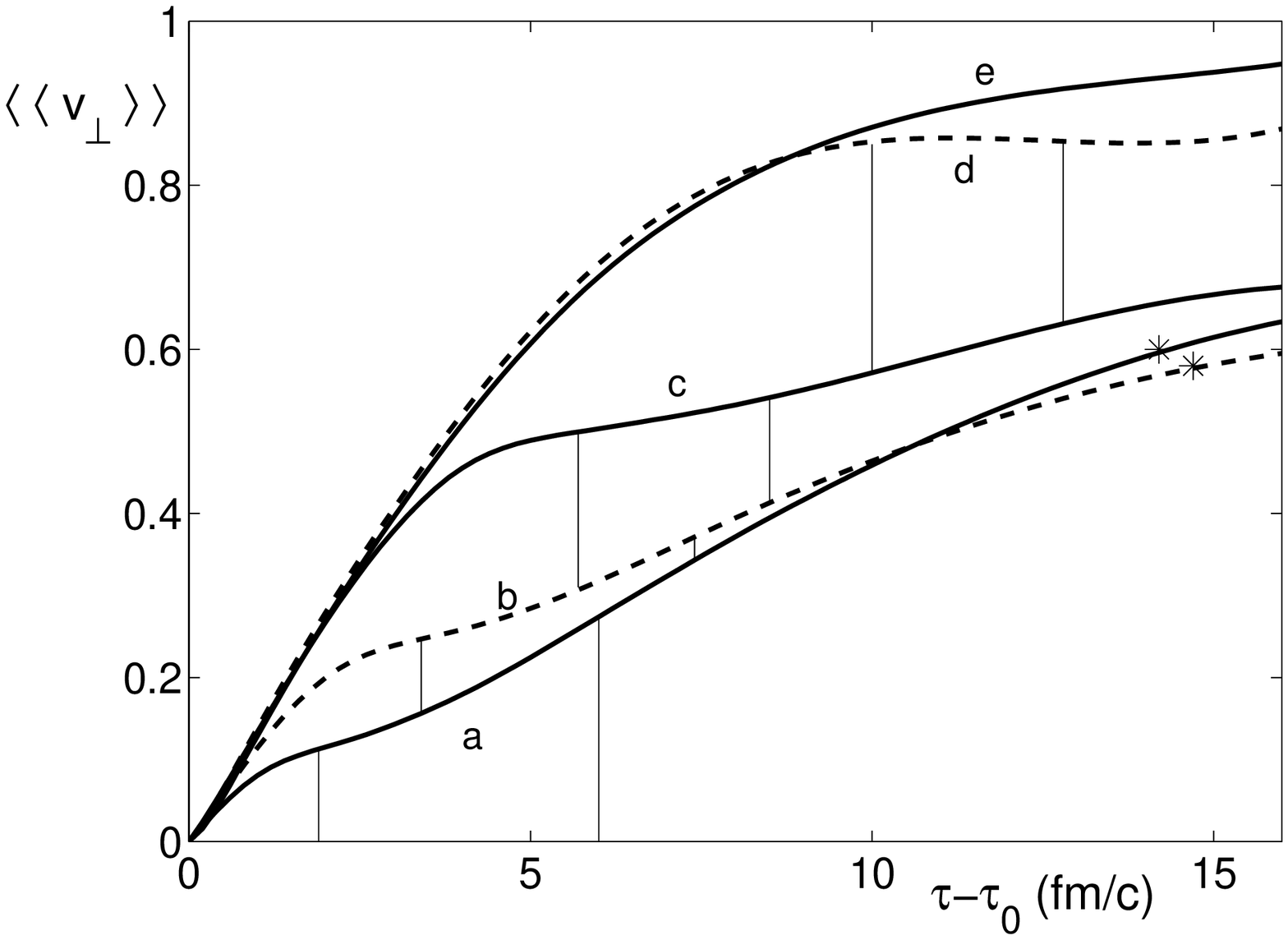,height=4.9cm,width=7.5cm}\\
\caption{Time evolution of the spatial ellipticity $\epsilon_x$, the 
  momentum anisotropy $\epsilon_p$, and the radial flow
  $\lda v_\perp \rda$. The labels {\tt a, b, c} and {\tt d} denote 
  systems with initial energy densities of 9, 25, 175 and 25000 
  GeV/fm$^3$, respectively, expanding under the influence of EOS~Q. 
  Curves {\tt e} show the limiting behaviour for EOS~I as $e_0\to\infty$ 
  (see text). In the lower two panels the two vertical lines below each 
  of the curves {\tt a-d} limit the time interval during which the 
  fireball center is in the mixed phase. In the upper panel the dots
  (crosses) indicate the time at which the center of the reaction zone 
  passes from the QGP to the mixed phase (from the mixed to the HG
  phase). For curves {\tt a} and {\tt b} the stars indicate the
  freeze-out point; for curves {\tt c-e} freeze-out happens outside
  the diagram. 
 \label{F7}}
\ece
\end{figure}

Figure \ref{F6} shows the analogous situation for EOS~Q (which includes
a phase transition) for identical initial conditions. Compared to 
Fig.~\ref{F5} one sees clear differences: the lack of a pressure
gradient in the mixed phase inhibits its transverse expansion; the 
hadronic phase ouside the mixed phase expands quickly and freezes out,
leaving a shell of mixed phase matter behind which inertially confines
the QGP matter in the center. The matter with the softest EOS (smallest
$p/e$) is concentrated around the QGP/mixed interface (thick contour at 
1.6 GeV/fm$^3$). When the QGP matter finally pushes the mixed phase 
shell apart (the ``nutcracker phenomenon'' discovered in \cite{TS99}),
the energy density contours develop an interesting structure vaguely
reminiscent of two separated half shells. Compared to Fig.~\ref{F5}, 
the elliptic flow clearly needs more time to push the matter from 
a state of positive to one of negative elliptic deformation. This is 
due to the inertia of the mixed phase shell which does not participate 
in the pushing.  

Figures \ref{F5} and \ref{F6} emphasize the spatial structure of
the fireball at fixed time steps. Let us now study the time evolution
in more detail. To this end we condense the information contained
in the density and flow patterns into three time-dependent scalar 
quantities: 

{\em (i)} The ``spatial ellipticity'' 
 \beq{eps_x}
   \epsilon_x = {\lda y^2{-}x^2 \rda \over \lda y^2{+}x^2 \rda}
 \eeq
characterizes the spatial deformation of the fireball in the transverse
plane. The angular brackets denote energy density weighted spatial 
averages at a fixed time. $\epsilon_x$ causes azimuthal anisotropies 
in the transverse pressure gradients which would eventually drive it 
to zero if the hydrodynamic evolution were not cut short by the 
freeze-out process.

{\em (ii)} The momentum anisotropy 
 \beq{eps_p}
   \epsilon_p = {\lda T^{xx}{-}T^{yy} \rda \over  
                 \lda T^{xx}{+}T^{yy} \rda} 
 \eeq   
measures in an analogous way the anisotropy of the transverse 
momentum-space density. It is directly calculated from the spatial 
components of the energy momentum tensor but, as shown in \cite{KSH99},
at freeze-out it is nearly equal to the $p_{_{\rm T}}^2$-weighted 
elliptic flow $v_{2,p_{\rm T}^2}$ for pions as calculated from their
final momentum spectra \cite{fn2}. Its time-dependence thus provides a 
picture of the dynamical buildup of the elliptic flow even at early 
times when the elliptic flow coefficient $v_2$ (which is calculated
from hadronic momentum spectra, see Sec.~\ref{sec4}) is not yet
defined. For pions at freeze-out $v_2$ is given by $2\,v_2 \approx$
$v_{2,p_{\rm T}^2} \approx \epsilon_p$ \cite{KSH99}. 

{\em (iii)} The time-dependence of the average radial flow velocity
 \beq{vperp}
   \lda v_\perp \rda = {\llda \gamma {\textstyle\sqrt{v_x^2{+}v_y^2}}\rrda 
                        \over \lda \gamma \rda} \,.
 \eeq
characterizes the buildup of the overall transverse expansion which is
modulated by the elliptic flow. Comparing the time-dependencies of
$\lda v_\perp \rda$ and $\epsilon_p$ allows to answer the question
to which stages of the expansion (i.e. to which domains of the EOS) 
each one is most sensitive.

\vspace*{-0.4cm}

We now give a detailed discussion of Figs.~\ref{F7}a-c which show 
(using EOS~Q) the time evolution for the above three quantities for 
a sequence of collision energies, parametrized by the initial central 
energy density in $b{\,=\,}0$ Pb+Pb collisions, $e_0$: $e_0$=\,9, 25, 
175, and 25000 GeV/fm$^3$ (curves {\tt a} through {\tt d} in 
Figs.~\ref{F7}). With increasing $e_0$ the initial time $\tau_0$ 
was scaled down as described at the end of Sec.~\ref{sec2e}. The 
lowest of these $e_0$-values corresponds to 158\,$A$\,GeV Pb+Pb 
collisions at the SPS, while the highest value is far beyond the 
reach of even the LHC.

A calculation with EOS~I is shown for comparison as curve {\tt e}. 
Since EOS~I ($e=3p$) is completely scale invariant, the time 
evolution of the dimensionless ratios (\ref{eps_x}), (\ref{eps_p}), 
and (\ref{vperp}) is invariant under a rescaling of $e_0$ as long 
as $\tau_0$ is held fixed (see Eqs.~(\ref{DGL})). Changing 
$\tau_0\sim e_0^{-1/4}$ breaks this scaling, but only weakly
as we have checked. Curves {\tt e} in Fig.~\ref{F7} show the time
evolution for EOS~I in the limit $e_0\to\infty,\ \tau_0\to 0$. Not 
shown is a calculation with EOS~Q which was initialized with an 
extraordinarily high initial temperature of $T_0\approx 20$\,GeV 
($e_0{\,=\,}25\times10^6$\,GeV/fm$^3$); during the first 16 fm/$c$ 
covered by Fig.~\ref{F7} it fully coincides with curve {\tt e}. In 
this case almost all of the matter stays in the QGP phase during this 
time period whose EOS coincides with EOS~I up to the (here negligible) 
bag constant. Therefore, as expected, the hydrodynamic evolution with 
EOS~Q approaches at asymptotically high energies that with EOS~I.

Inspection of Fig.~\ref{F7} shows that the elliptic flow $\epsilon_p$
saturates at large times while the radial flow $\lda v_\perp \rda$
keeps rising forever, albeit at a decreasing rate. The driving force 
for radial flow, the radial pressure gradient between the matter in
the fireball and the surrounding vacu\-um, never vanishes completely. 
The spatial ellipticity $\epsilon_x$, on the other hand, which is 
responsible for azimuthal anisotropies in the transverse pressure 
gradients and thus drives the evolution of $\epsilon_p$, passes 
through zero after some time. Afterwards the longer axis of the 
transverse fireball cross section no longer points perpendicular to 
the reaction plane, but {\em into} the reaction plane. 
A vanishing $\epsilon_x$ implies a vanishing growth rate for 
$\epsilon_p$; as $\epsilon_x$ turns negative, smaller oppositely 
directed anisotropies of the pressure gradients develop which can 
actually cause $\epsilon_p$ to decrease again. This can be seen in 
Fig.~\ref{F7}b for large values of $e_0$ where the sign of 
$\epsilon_x$ changes sufficiently early in the collision that 
pressures are still high enough to generate this effect. 

Qualitatively one hence can say that the final value of $\epsilon_p$ is 
established roughly at the point when $\epsilon_x$ passes through 
zero. For SPS energies this happens just before decoupling (implying
that the fireball freezes out in a nearly circular configuration), but
at high energies this occurs well before freeze-out. Generically the
freeze-out value of $\epsilon_p$ (and thus $v_2$) is sensitive to the 
EOS at significantly higher energy densities than the radial flow 
$\lda v_\perp \rda$. {\em The elliptic flow indeed measures the early
pressure} \cite{S97,S99}.
 
On a more detailed level, the time evolution shows an interesting 
additional feature: In curves {\tt b} and {\tt c} the elliptic flow 
$\epsilon_p$ is seen to peak even {\em before} $\epsilon_x$ passes 
through zero. The origin of this phenomenon, which is related to the
phase transition, will be discussed in Sec.~\ref{sec3c} below.

Comparison of the lower two panels in Fig.~\ref{F7} shows that the
softening effect on the EOS of the phase transition affects the
buildup of $\lda v_\perp \rda$ and $\epsilon_p$ at similar times.
However, the influence on $\epsilon_p$ is stronger since elliptic 
flow is a smaller effect (which feels only the anisotropies in
the transverse pressure gradient, not its overall magnitude) and thus
more fragile than radial flow. This results in a relatively larger 
sensitivity of elliptic flow to the phase transition.

\suse{Central U+U collisions in the side-on-side configuration}
\label{sec3b}

As discussed in Sec.~\ref{sec2c}, central U+U collisions in the
side-on-side configuration provide 14\% higher energy density over
nearly twice the volume at the same initial spatial deformation as
Pb+Pb collisions at $b{\,=\,}7$\,fm. This leads to a longer lifetime
for non-zero spatial ellipticity $\epsilon_x$, the driving force for 
elliptic flow, and also for the whole fireball until freeze-out. 
Hence the system has more time for thermalization, favoring the 
applicability of our hydrodynamic method. For this reason we decided 
to perform quantitative calculations for this system and make
predictions for experiments with uranium beams at RHIC and LHC.  

\begin{figure}[htbp]
\bce
\epsfig{file=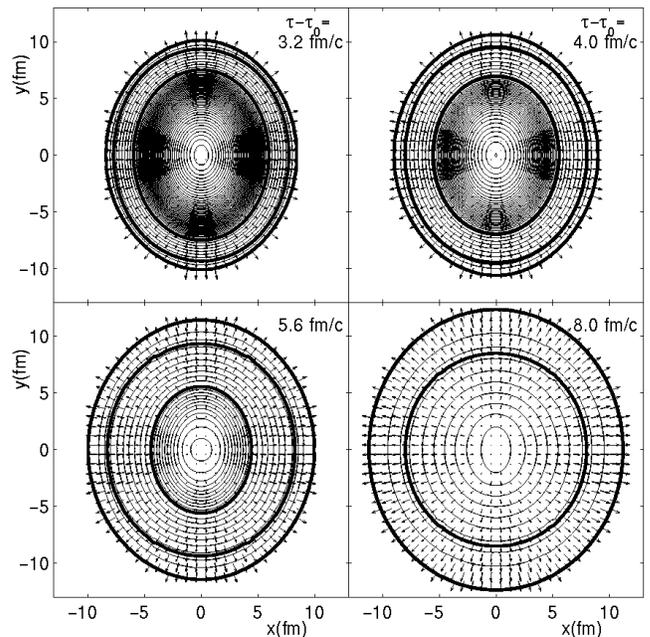,width=8.5cm}
\caption{Same as Figs.~\ref{F5} and \ref{F6} ($e_0{\,=\,}175$\,GeV/fm$^3$ 
  at $\tau_0{\,=\,}0.38$\,fm/$c$, EOS~Q), but for central side-on-side 
  U+U collisions. The spacing between energy density contours is again 
  150 MeV/fm$^3$, and the snapshots are taken at the same times. The 
  corresponding maximum energy densities are 8.71, 6.06, 3.27, and
  1.47\,GeV/fm$^3$, respectively. See text for discussion.
\label{F8}}
\ece
\end{figure}
\vspace*{-0.5cm}

We first look once more at the space-time evolution of the transverse 
energy density and flow profiles, shown in Fig.~\ref{F8}. The
initialization corresponds to the same collision energy as in 
Fig.~\ref{F6} ($e_0{\,=\,}175$ GeV/fm$^3$) but, since we now consider
central ($b{\,=\,}0$) collisions, the initial energy density in the
center of the deformed collision\break
%
\begin{figure}[ht]
\bce
\hspace*{0.2cm}\epsfig{file=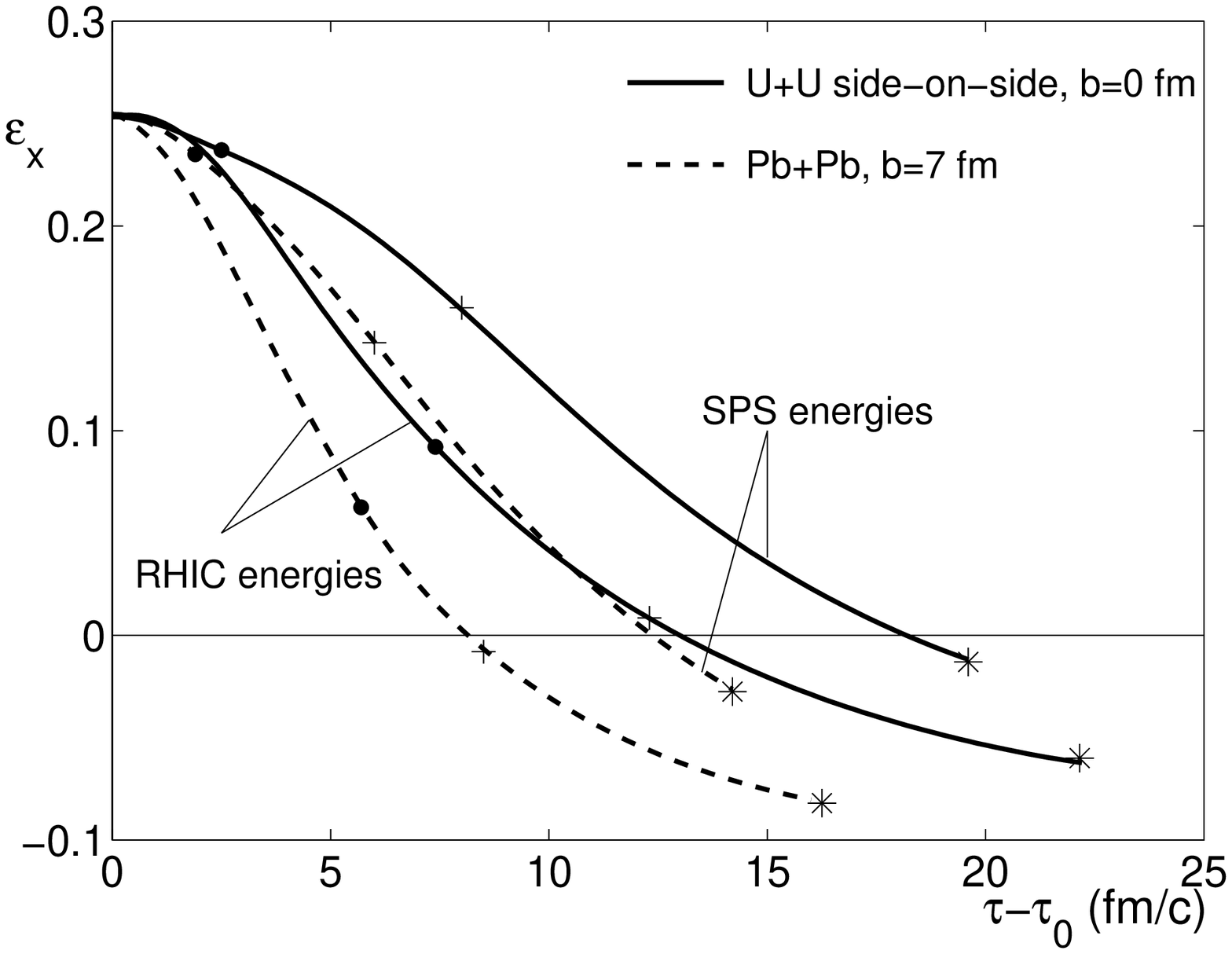,height=5cm,width=7.3cm}\\
\epsfig{file=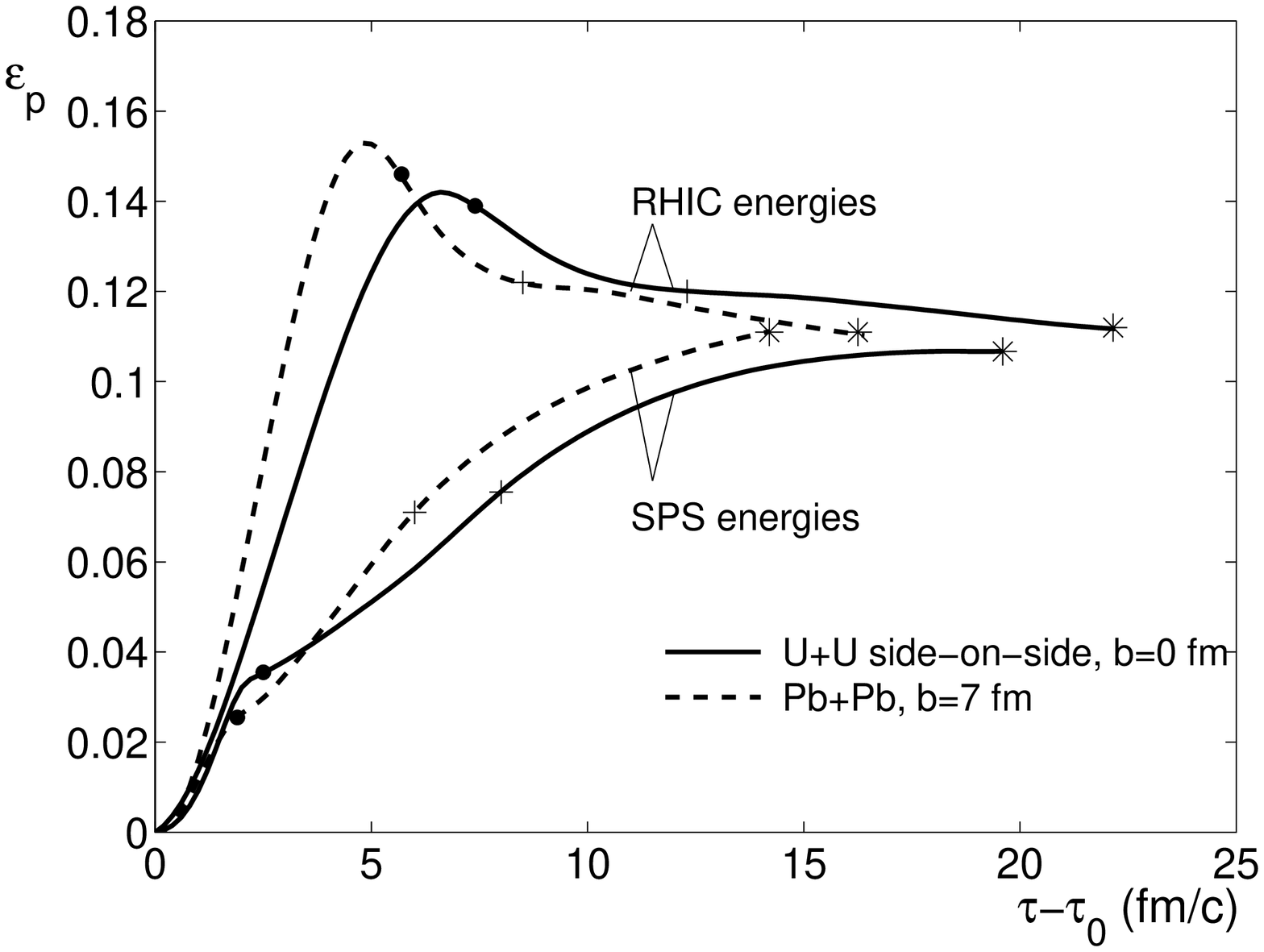,height=5cm,width=7.5cm}\\
\hspace*{-0.2cm}\epsfig{file=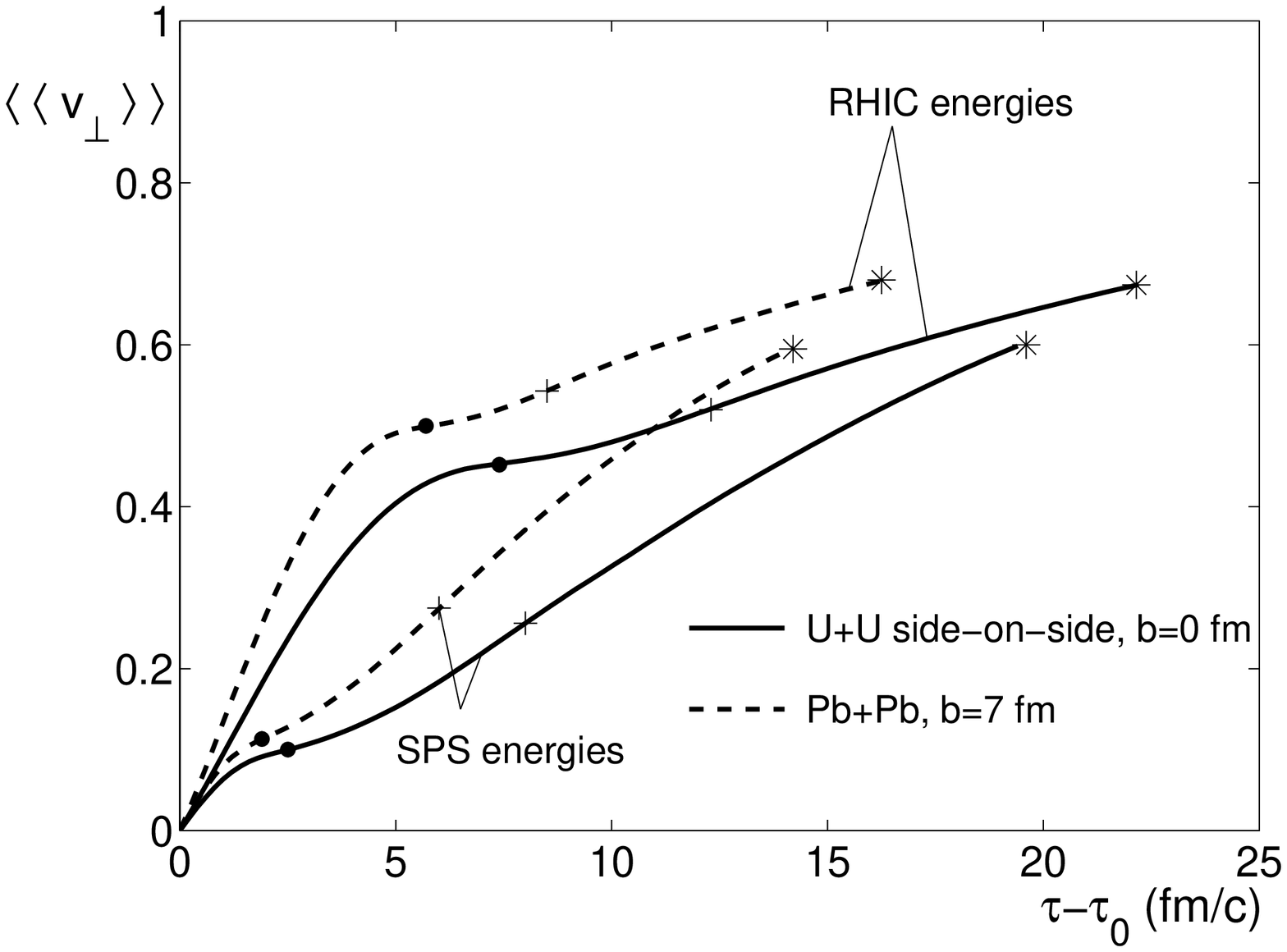,height=5cm,width=7.6cm}\\
\caption{Same as Fig.~\ref{F7}, but now comparing central U+U (solid) 
  to semiperipheral ($b{\,=\,}7$\,fm) Pb+Pb collisions (dashed) at
  two selected beam energies. The curves labelled ``SPS'' correspond 
  to $e_0$\,=\,9\,GeV/fm$^3$ (8.3\,GeV/fm$^3$) for central Pb+Pb
  (side-on-side U+U) collisions, those labelled ``RHIC'' have 
  $e_0$\,=\,175\,GeV/fm$^3$ in both cases.
 \label{F9}}
\ece
\end{figure}
%
\noindent
region is higher than in the
semiperipheral Pb+Pb collisions of Fig.~\ref{F6}. As seen in 
Fig.~\ref{F9}, the whole time evolution is slower for central U+U 
than for semiperi\-phe\-ral Pb+Pb collisions, due to the larger system 
size: At $\tau-\tau_0{\,=\,}3.2$\,fm/$c$ (the first shown snapshot)
the central energy density is 50\% higher, and at
$\tau-\tau_0{\,=\,}8$\,fm/$c$ (the last snapshot) it is even by a
factor 3 larger than in $b{\,=\,}7$\,fm Pb+Pb collisions at the same
beam energy. Freeze-out occurs nearly 30\% later in central U+U than
in semiperipheral Pb+Pb collisions (see Fig.~\ref{F9}).

We note with surprise that the ``nutcracker'' pheno\-me\-non \cite{TS99} 
is conspicuously missing in the U+U col\-li\-sions. We could not find
it at lower and higher collision energies either. The origin of this
difference between central U+U and peripheral Pb+Pb collisions will 
be discussed in the following subsection.

In Fig.~\ref{F9} we compare the time evolutions of the three 
characteristic quantities $\epsilon_x,\,\epsilon_p$, and 
$\lda v_\perp\rda$ in central U+U and semiperipheral Pb+Pb 
collisions, at SPS ($e_0$\,=\,9\,GeV/fm$^3$) and RHIC 
($e_0{\,=\,}175$\,GeV/fm$^3$) energies. We note that at freeze-out
($T_{\rm dec}{\,=\,}120$\, MeV) both systems give nearly the same 
radial and elliptic flow, in spite of the different time 
evolution: in the large system both flow types develop more slowly, 
but over a longer time. This does not take into account that the
flow gradients are smaller in the larger system, leading to later
freeze-out at a lower temperature \cite{HS98}. This would not
change the elliptic flow since $\epsilon_p$ has already saturated 
(actually, it would lead to a very slight decrease of $\epsilon_p$, 
see Fig.~\ref{F9}b). The radial flow $\lda v_\perp \rda$ would, 
however, be somewhat larger. Since we enforced freeze-out at the 
same value $T_{\rm dec}$, we don't see this.

\suse{What makes the nut crack?}
\label{sec3c}

In this subsection we analyze two questions which so far remained 
open: (1) Why does the ``nutcracker'' phenomenon arise in 
semiperipheral Pb+Pb collisions, but not in central U+U collisions, 
in spite of their identical initial deformation? (2) What is the 
origin of the decrease of $\epsilon_p(\tau)$ before $\epsilon_x$ 
passes through zero which is observed in Fig.~\ref{F9}b and curves 
{\tt b} and {\tt c} of Fig.~\ref{F7}b?

\begin{figure}[ht]
\bce
\epsfig{file=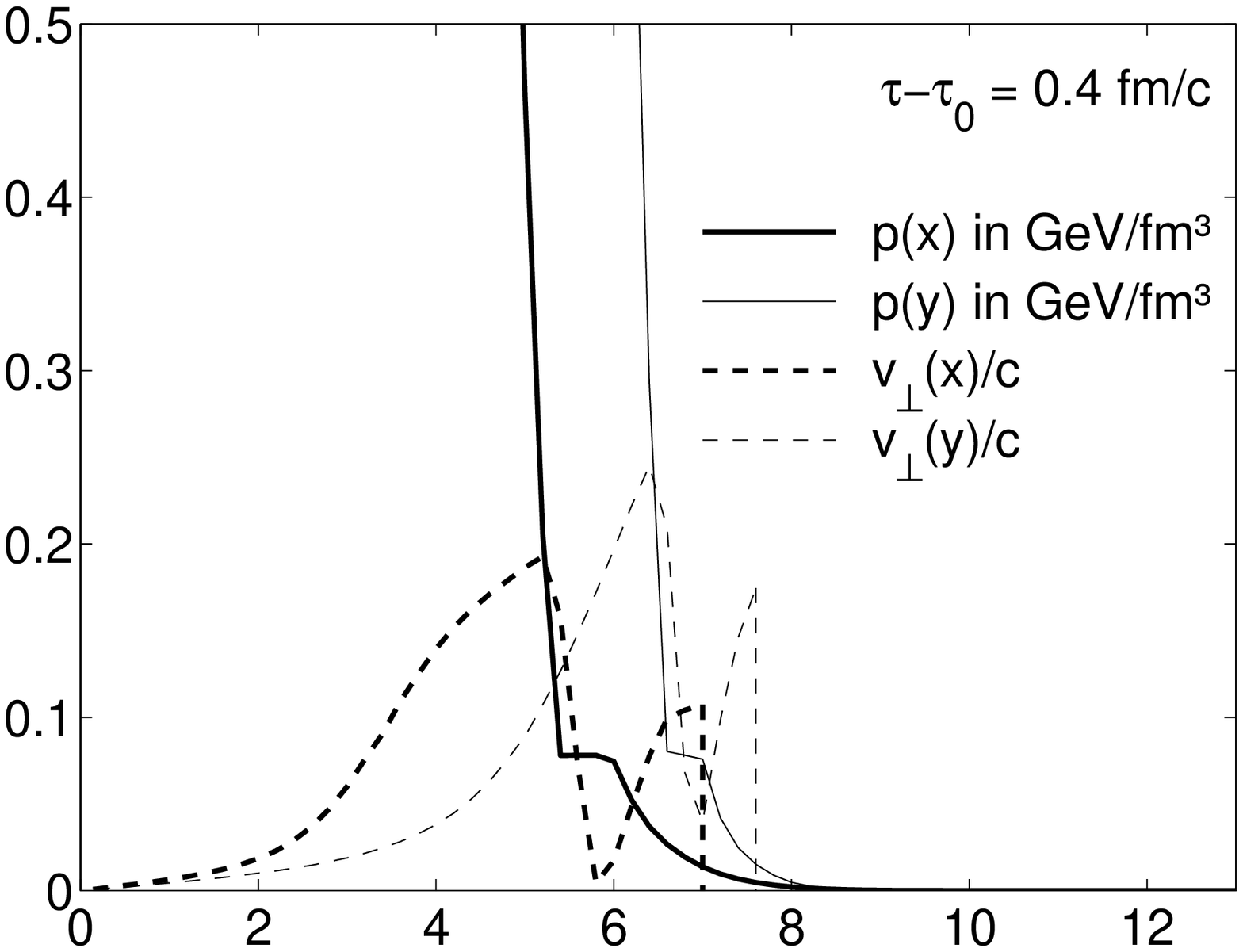,width=4.25cm}
\epsfig{file=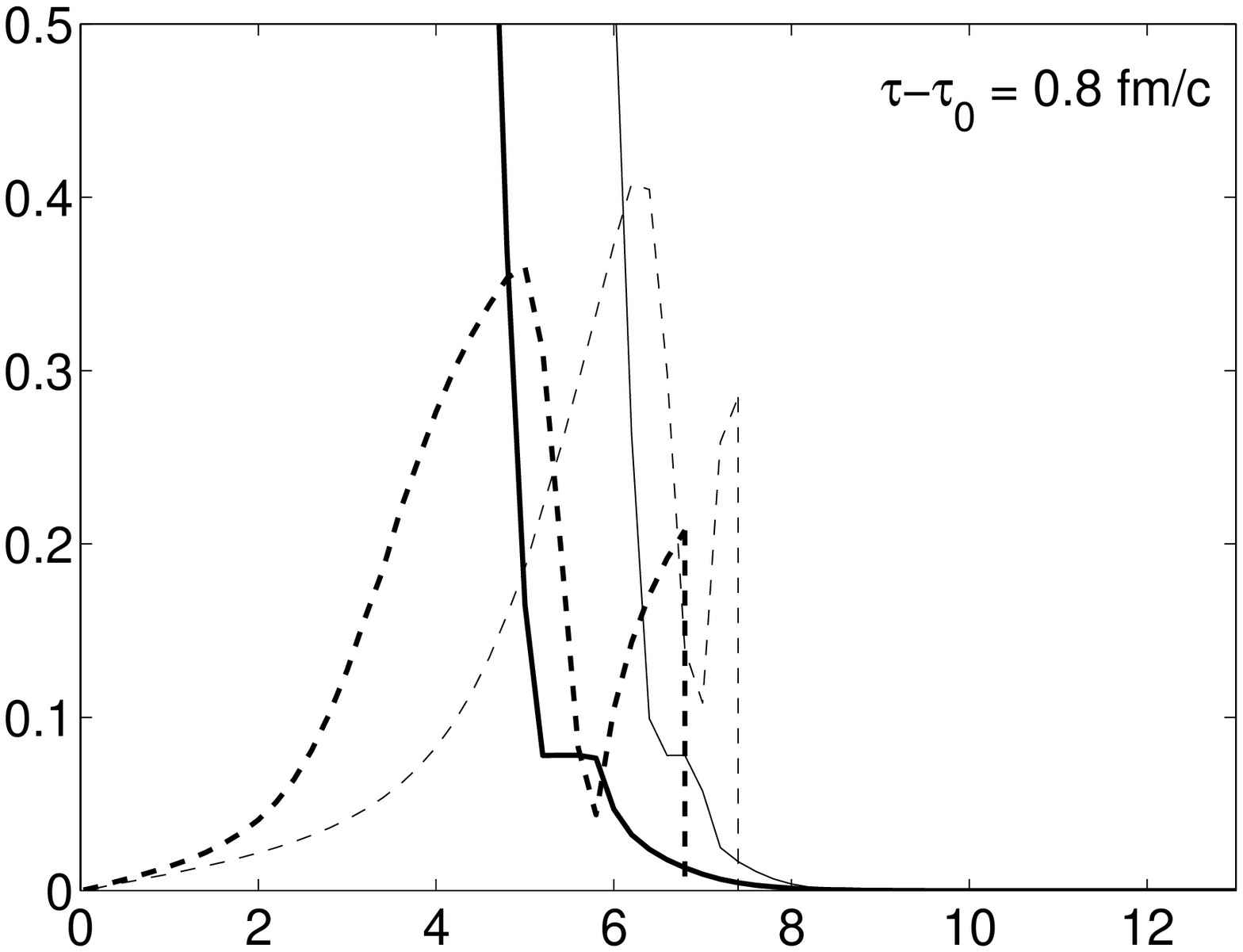,width=4.25cm}
\epsfig{file=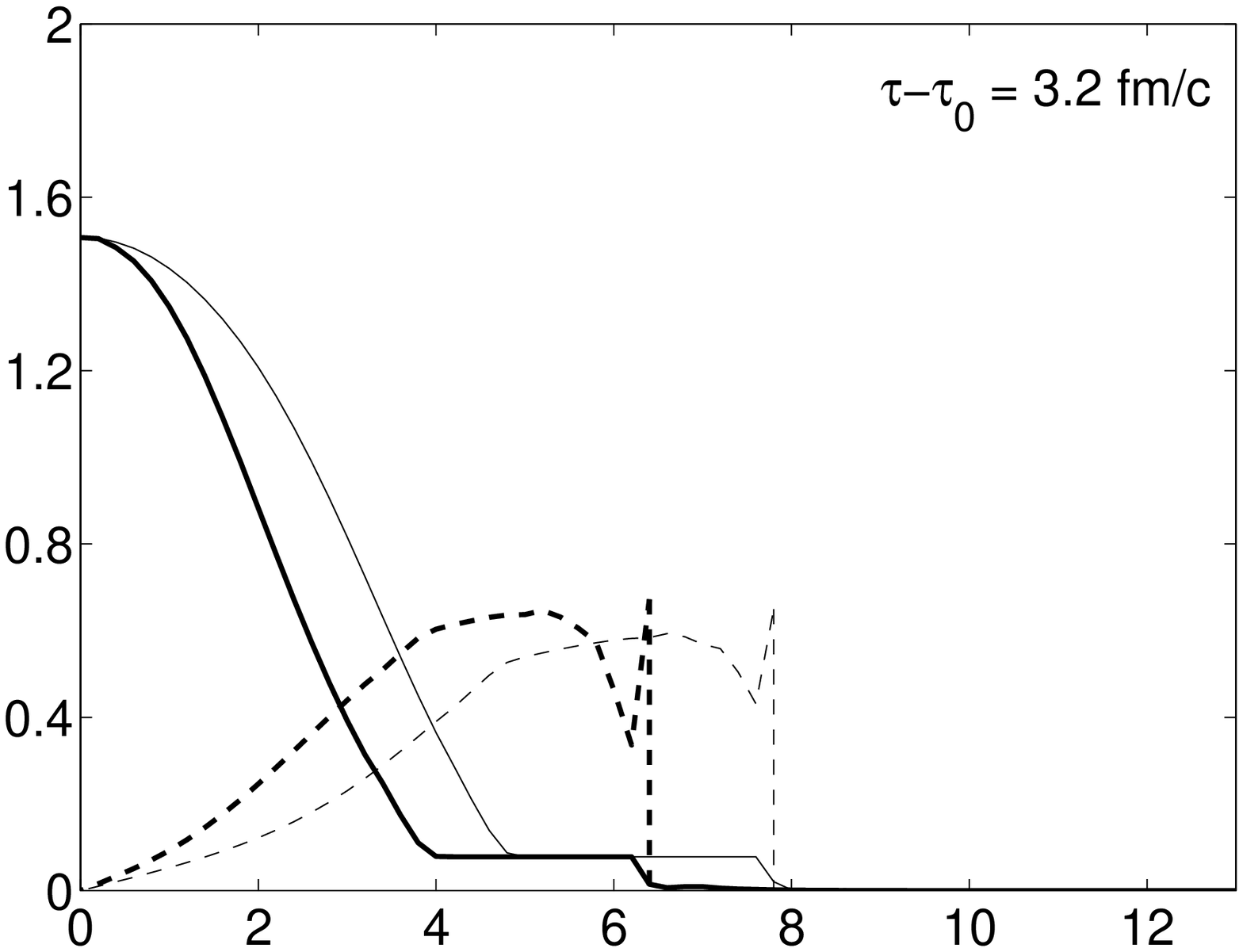,width=4.25cm}
\epsfig{file=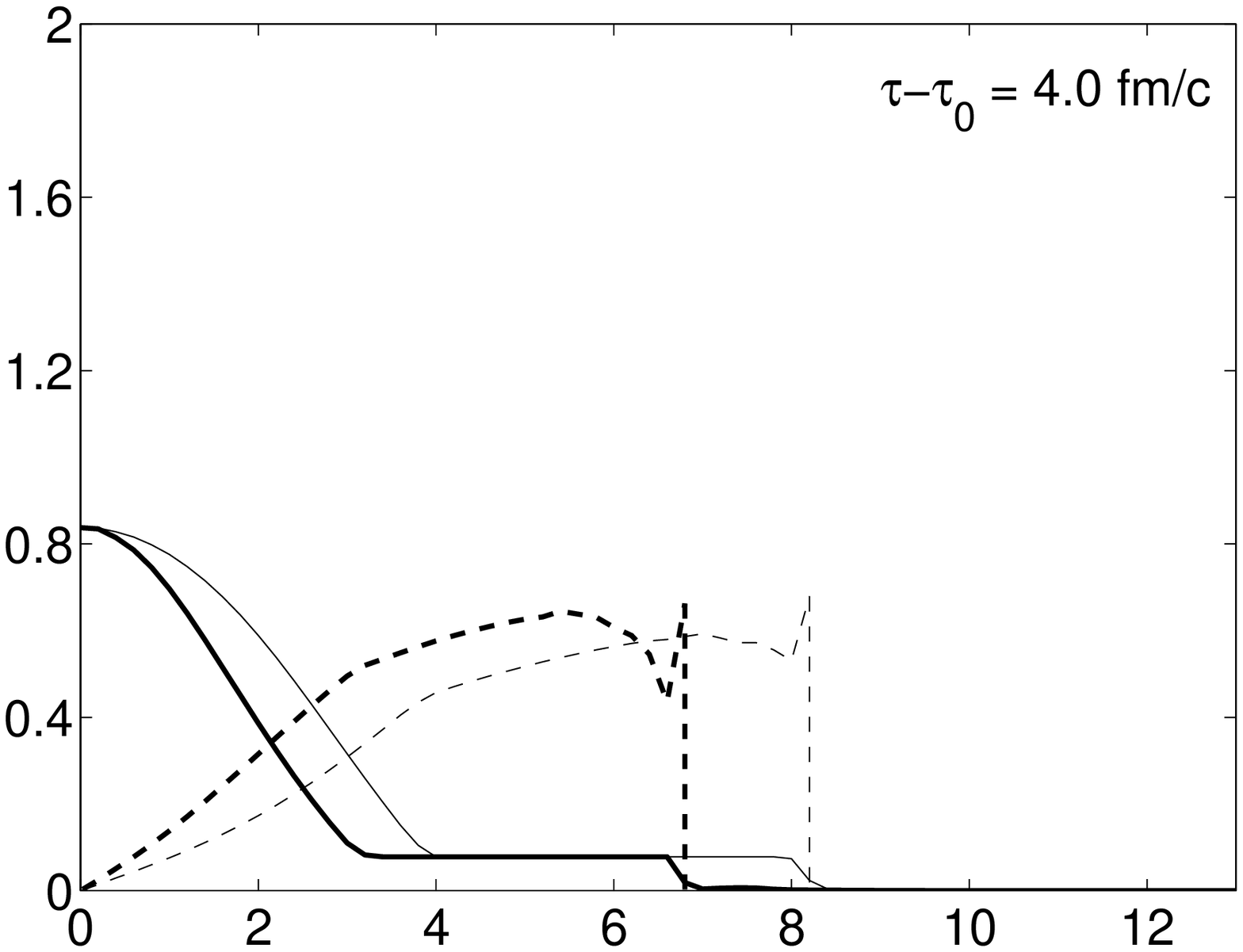,width=4.25cm}
\epsfig{file=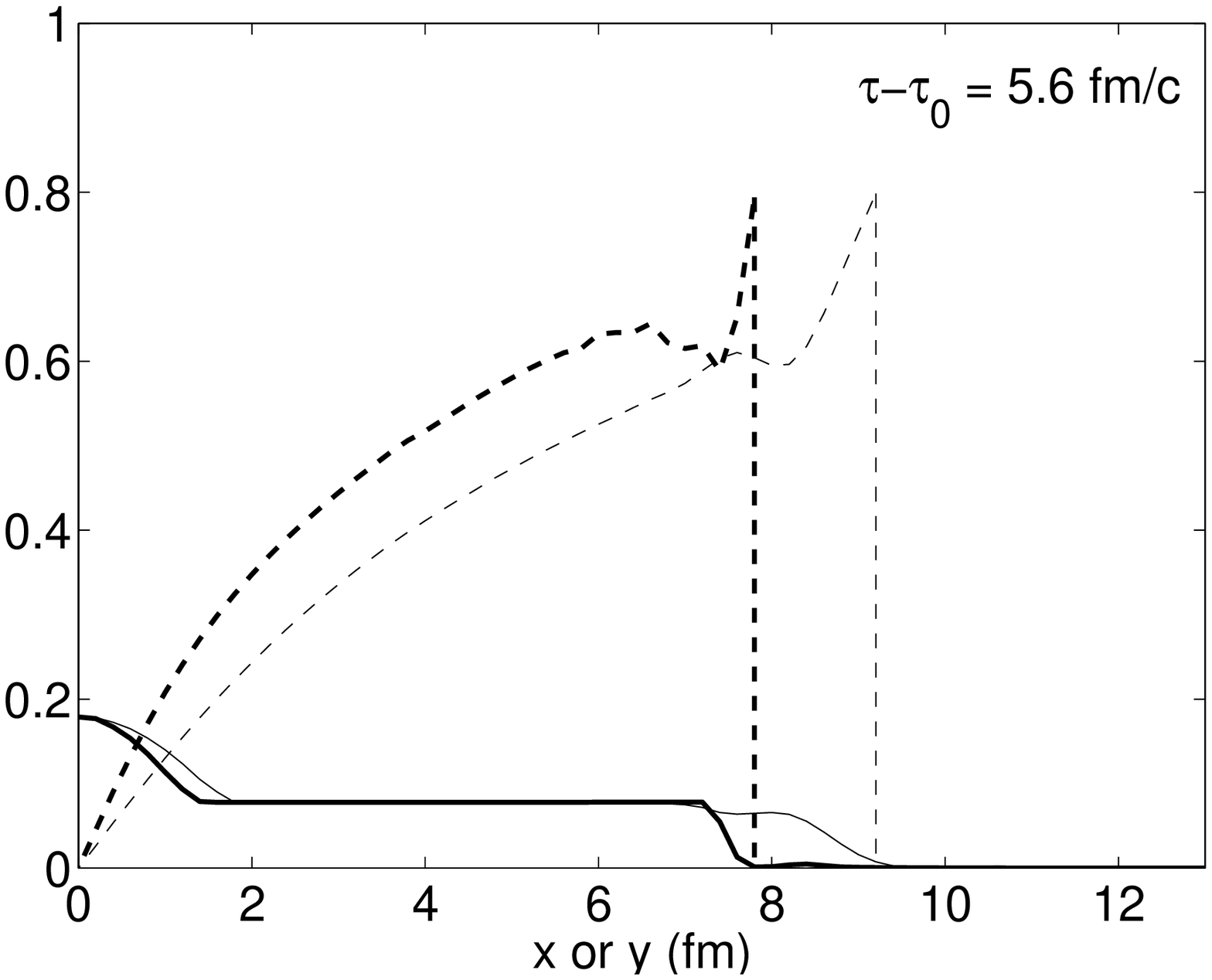,width=4.25cm}
\epsfig{file=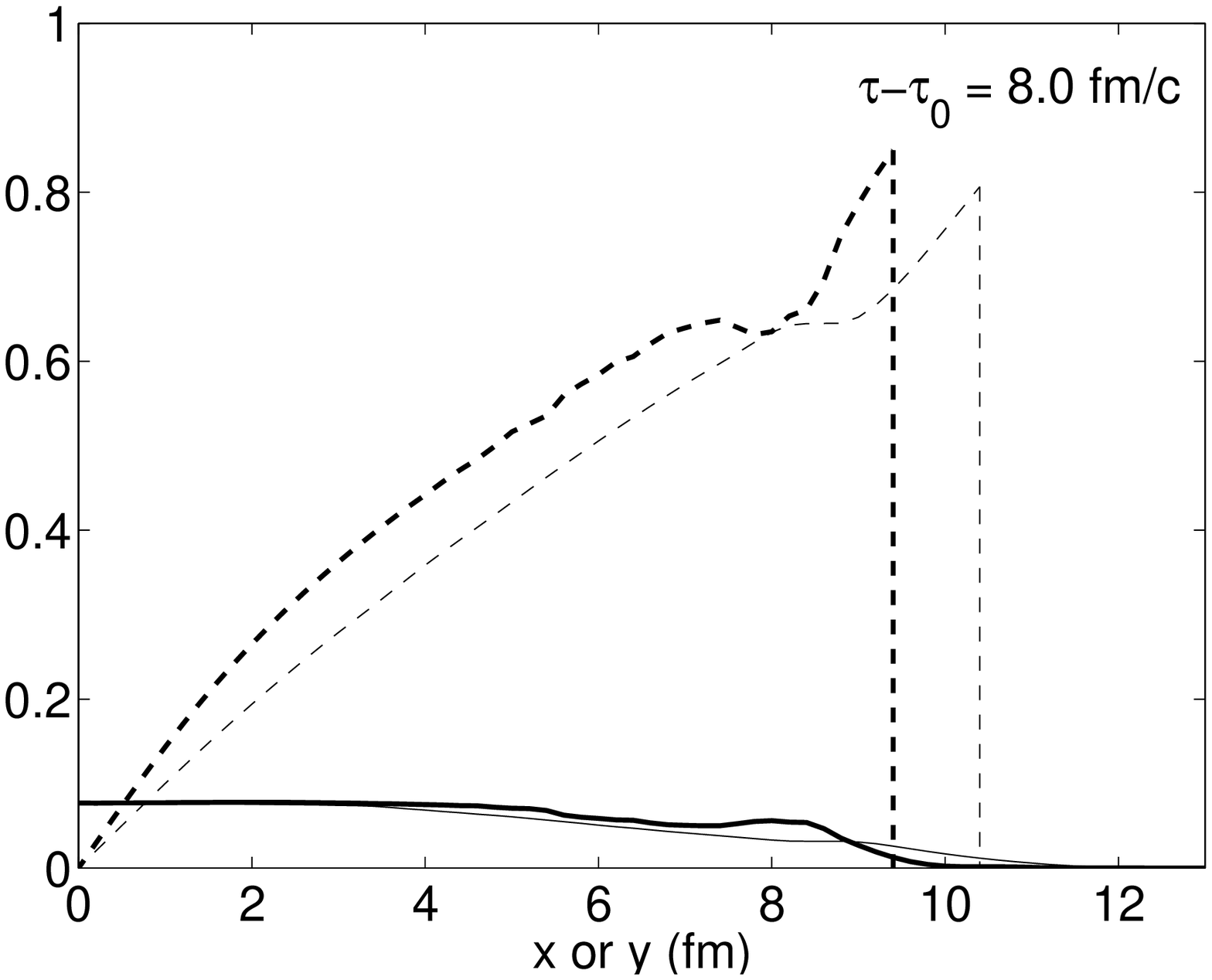,width=4.25cm}\\
\vspace*{0.2cm}
\caption{Transverse pressure (solid) and velocity (dashed) profile,
  in $x$ (thick) and $y$ (thin) directions, for Pb+Pb collisions
  at $b{\,=\,}7$\,fm. The 6 panels show snapshots at the indicated times.
  The region of nearly constant pressure is in the mixed phase. The 
  velocity profiles (dashed) are cut off at the freeze-out point. Initial
  conditions as in Fig.~\ref{F6}.
 \label{F10}}
\ece
\end{figure}

\begin{figure}[ht]
\bce
\epsfig{file=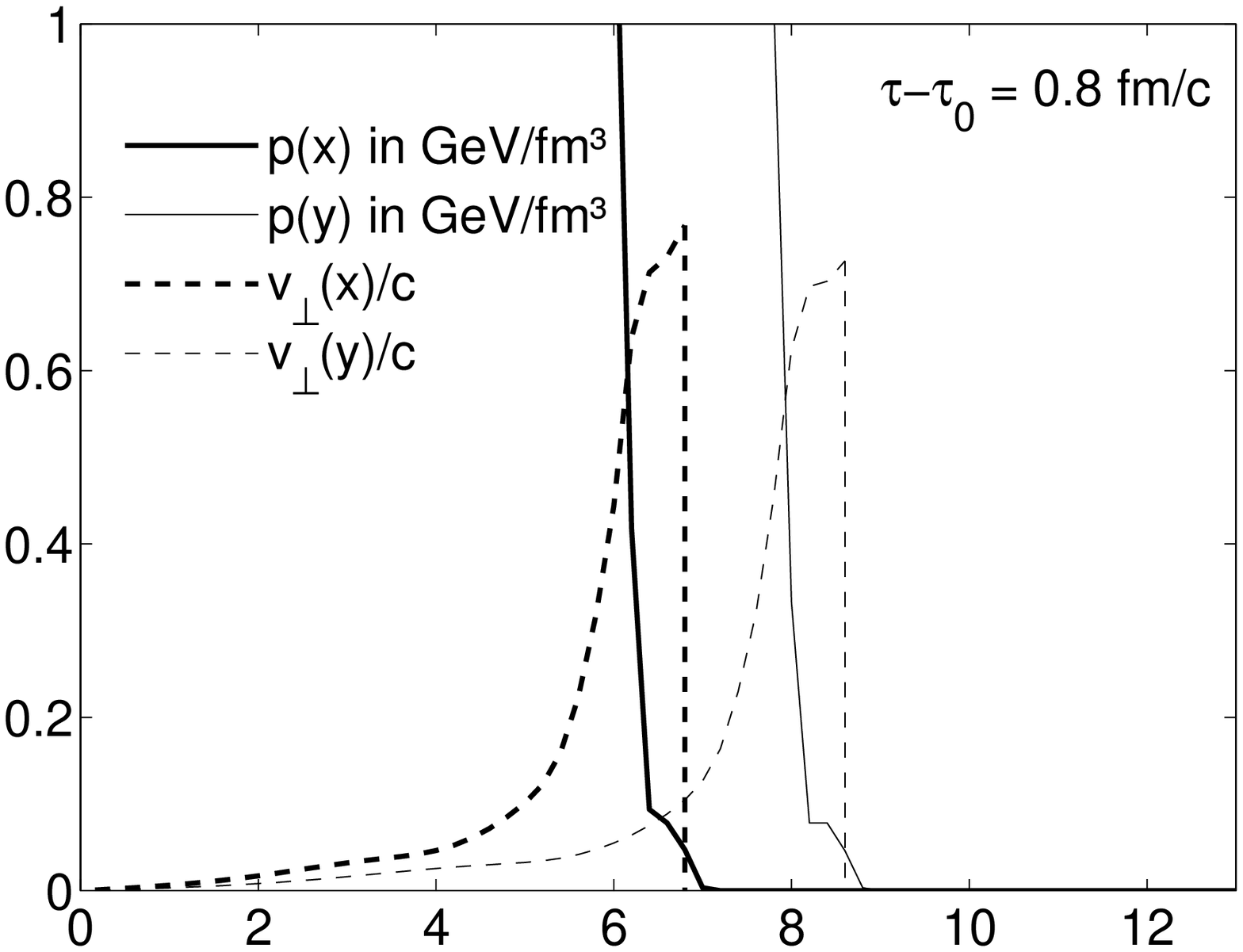,width=4.25cm}
\epsfig{file=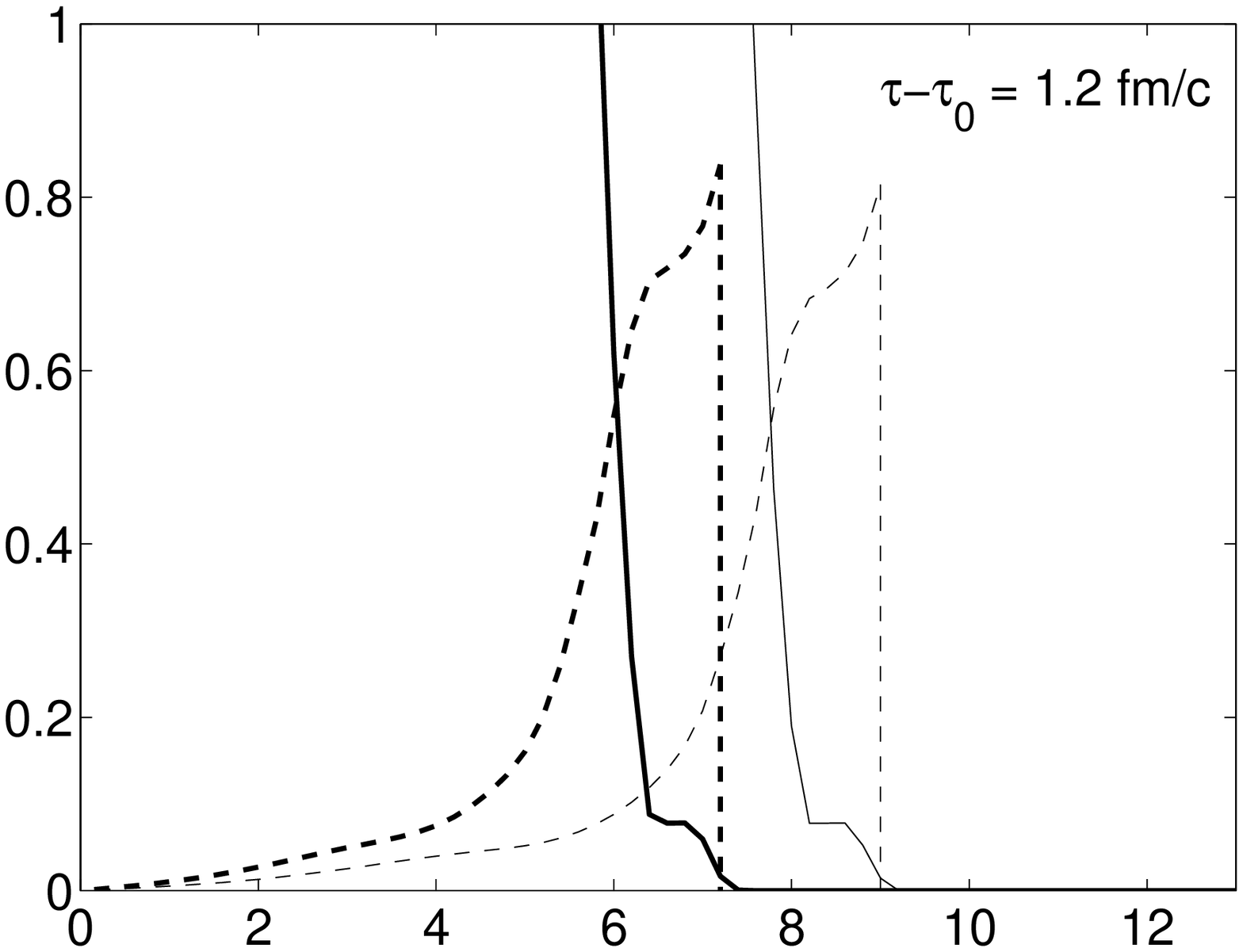,width=4.25cm}
\epsfig{file=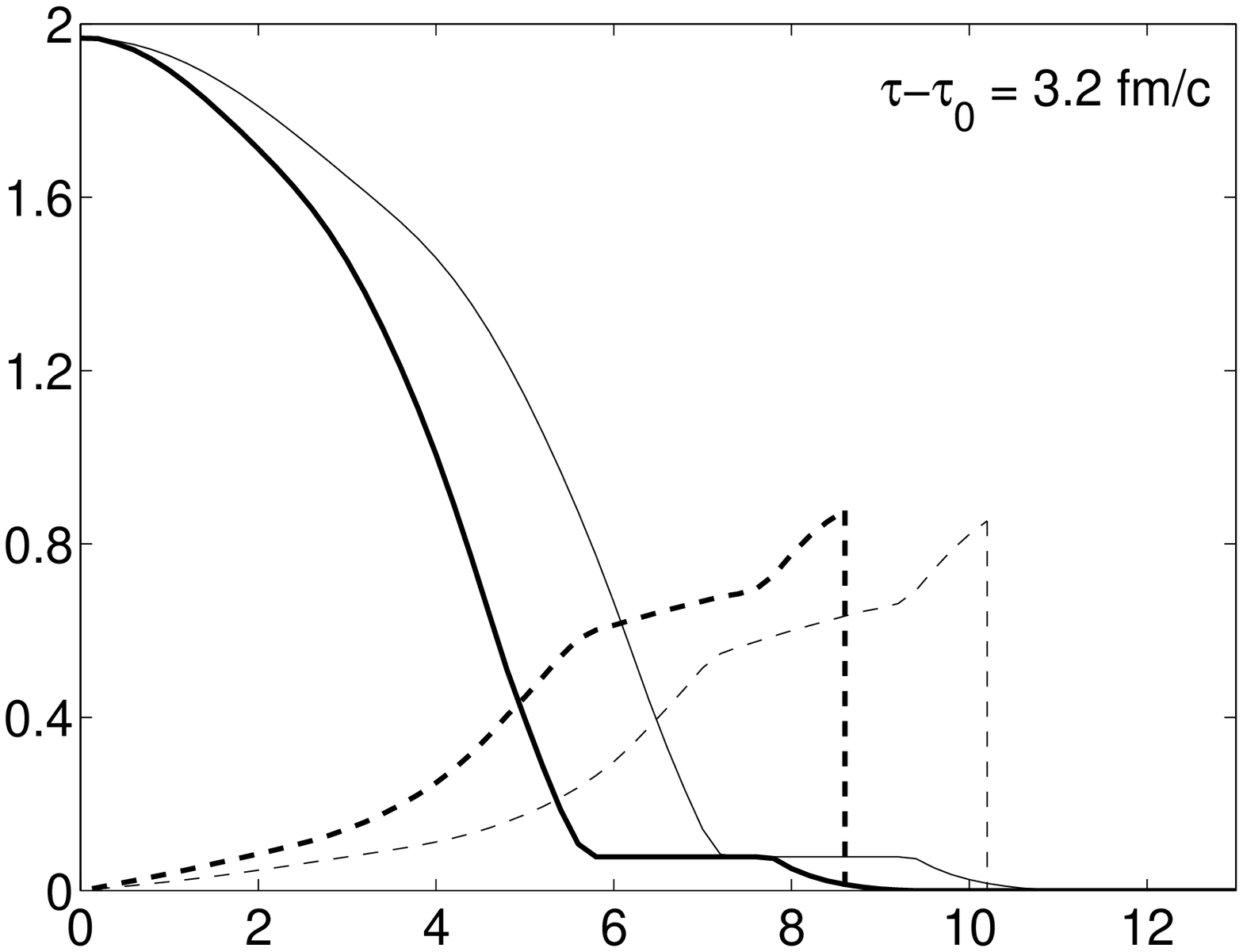,width=4.25cm}
\epsfig{file=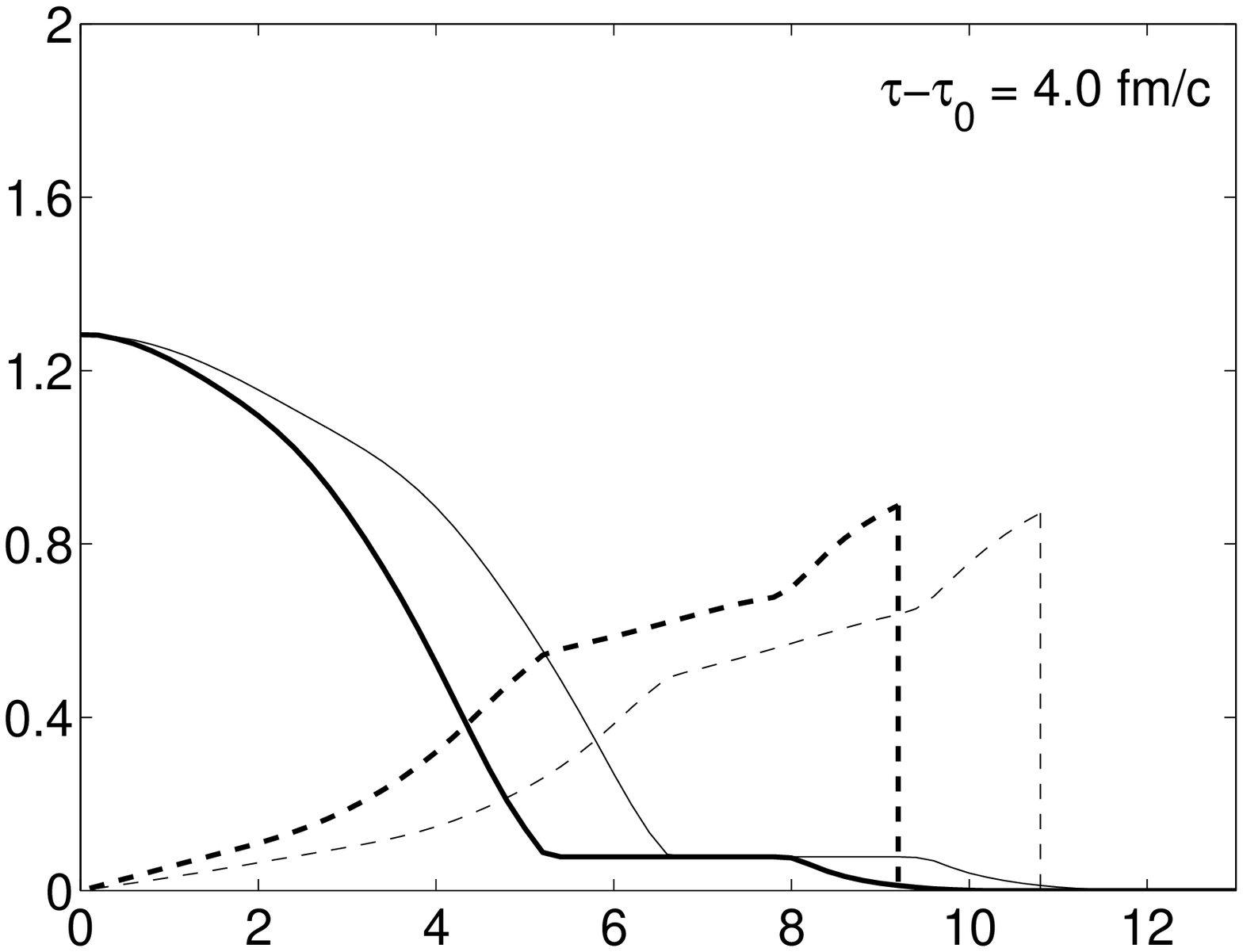,width=4.25cm}
\epsfig{file=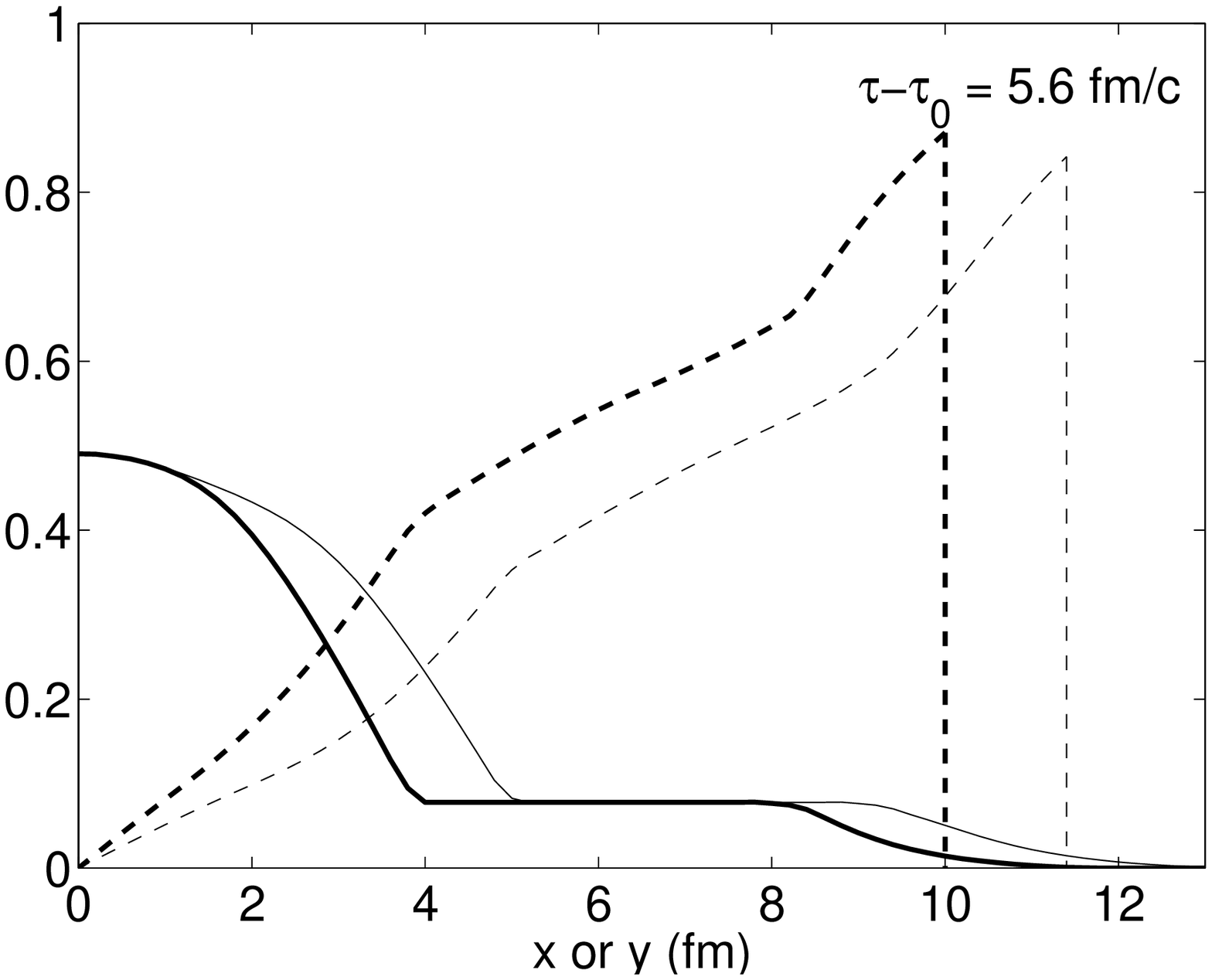,width=4.25cm}
\epsfig{file=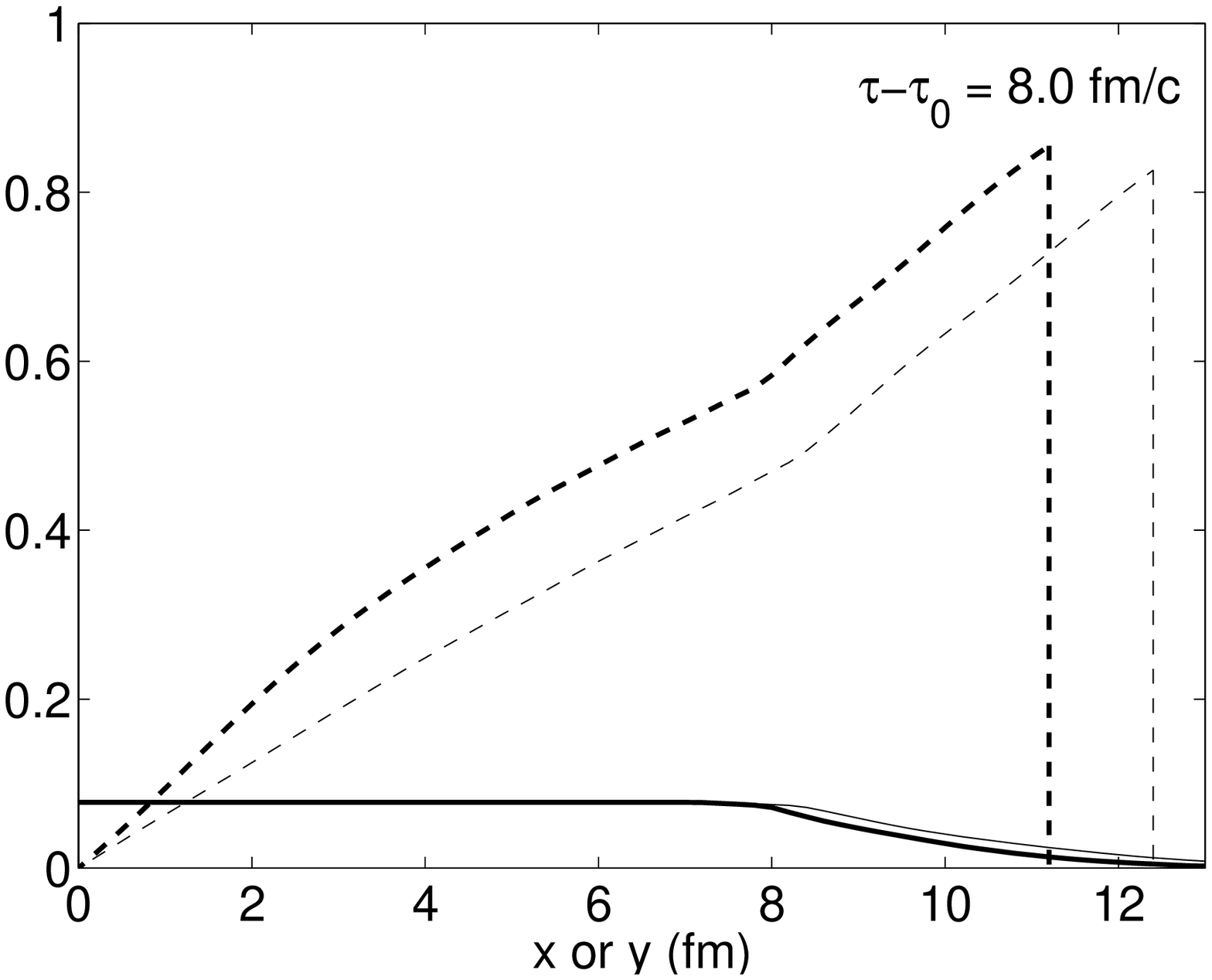,width=4.25cm}\\
\vspace*{0.2cm}
\caption{Same as Fig.~\ref{F10}, but for central U+U collisions
  in the side-on-side configuration. Initial conditions as in 
  Fig.~\ref{F8}.
 \label{F11}}
\ece
\end{figure}

To answer them requires a more detailed look at the time evolution
of the transverse pressure gradients (cause) and transverse flow 
profiles (effect). In Figs.~\ref{F10} and \ref{F11} we show a series
of six snapshots each for semiperipheral Pb+Pb and central U+U 
collisions, plotting the pressure and flow velocity profiles along 
the $x$ and $y$ axis, respectively. The crucial difference between the 
two collision systems is that in the semiperipheral Pb+Pb collisions
the initial fireball contains a roughly 0.5\,fm thick {\em layer of
  mixed phase matter with vanishing transverse flow velocity}; for
central U+U collisions the initial ener\-gy density drops to zero so
steeply that the mixed phase layer is initially practically absent.

As the matter begins to expand and dilute, a mixed phase layer begins
to develop also in the U+U collisions; however, due to the buildup of
transverse flow in the expanding matter, it is automatically created
with a {\em nonvanishing} transverse flow velocity. Thus, even without
pressure gradients inside the mixed phase which could accelerate it,
the mixed phase matter flows in the transverse directions, with
velocities exceeding those of the enclosed QGP matter (see
Fig.\,\ref{F11}). The resulting transverse flow profiles are 
monotonous functions of $x$ and $y$, with a selfsimilar (linear
``scaling'') pattern inside the mixed phase exactly as given by the
analytic solution recently found by Bir\'o \cite{Biro}. The monotony
of the transverse flow profiles is related one-to-one to the absence
of the nutcracker phenomenon.

In the semiperipheral Pb+Pb collisions, on the other hand, the
initially present mixed phase layer is at rest and, due to the lack of 
pressure gradients, cannot accelerate itself in the transverse
direction. As the transverse pressure gradients in the enclosed QGP
matter begin to accelerate the QGP matter, the latter ``slams'' into
the motionless mixed phase. This is clearly seen in the first four
panels of Fig.\,\ref{F10} which show a strong radial increase of
the transverse flow velocities inside the QGP phase, followed by a
dramatic drop inside the mixed phase and a second rise in the HG
matter near the edge. Inside the mixed phase the radial velocity
profile is thus completely different from the selfsimilar scaling
pattern seen in Fig.\,\ref{F11}. As time proceeds, this anomalous
structure in the Pb+Pb collisions weakens, and the velocity profile
begins to approach a scaling form inside the mixed phase; scaling 
violations survive longest near the outer edge of the mixed phase
layer. In the $y$ direction they disappear slightly earlier than
in the shorter $x$ direction; this is the origin of the ``nutcracker
phenomenon''. 

\vspace{-0.7cm}

Now we can also understand the decrease of $\epsilon_p$ even before
$\epsilon_x$ passes through zero: Figs.\,\ref{F7}, \ref{F10} and
\ref{F11} show that this happens while most of the fireball is in the
mixed phase. (Actually, $\epsilon_p$ begins to decrease while there is
still a small QGP core in the center.) During this stage the matter
expands essentially without transverse acceleration, featuring a 
nearly selfsimilar transverse flow pattern. While it lasts, the
selfsimilar flow dilutes the earlier developed momentum anisotropy
$\epsilon_p$. This feature is therefore also directly related to the
phase transition.   

\se{Experimental predictions}
\label{sec4}

While the time-evolution of $\epsilon_x$, $\epsilon_p$ and 
$\lda v_\perp \rda$ is interesting and helpful for an understanding
of the relevant physical mechanisms, only the final values
at freeze-out are observable (through the momentum spectra and, in 
the case of $\epsilon_x$, possibly indirectly via two-particle momentum 
correlations). The flow observables thus represent time-integrals over 
the expansion history and EOS, and their measurement in a single 
collision system at fixed beam energy provides very little information. 
Using flow signatures as indicators for properties of the equation of 
state for strongly interacting matter requires their measurement over a
wide range of external control parameters, such as impact parameter, size 
of the colliding nuclei, and beam energy. As discussed in the preceding
section, a time-differential measurement is to some extent possible by
comparing the radial and elliptic flow as functions of these parameters.

Flow anisotropies reflect themselves as non-vanishing higher order 
Fourier coefficients in a Fourier expansion of the azimuthal dependence
of the measured single-particle spectra around the beam direction 
\cite{VZ96}:
 \beq{vn}
   v_n(y) = {\int_{-\pi}^\pi d\phi\,\cos(n\phi)\, {dN \over dy\, d\phi}
             \over
             \int_{-\pi}^\pi d\phi\, {dN \over dy\, d\phi}} \,,\quad
   n=1,2,\dots  \,.
 \eeq
Since most experiments have limited $p_{_{\rm T}}$ acceptance, one 
studies these coefficients also as functions of the transverse momentum:
 \beq{vnpt}
   v_n(y,p_{_{\rm T}}) = 
   {\int_{-\pi}^\pi d\phi\,\cos(n\phi)\,
   {dN \over dy\,p_{\rm T}\,dp_{\rm T}\, d\phi}
   \over
   \int_{-\pi}^\pi d\phi\,
   {dN \over dy\,p_{\rm T}\,dp_{\rm T}\, d\phi}}
   \,.
 \eeq
The $p_{_{\rm T}}^2$-weighted anisotropic flow coefficients are
defined by
 \beq{vnpt2}
   v_{n,p_{\rm T}^2}(y) = 
   {\int_{-\pi}^\pi d\phi\,\cos(n\phi)\, \int p_{\rm T}^2\, dp_{\rm T}^2
    \, {dN \over dy\,dp_{\rm T}^2\, d\phi}
    \over
    \int_{-\pi}^\pi d\phi\,\int p_{\rm T}^2\, dp_{\rm T}^2
   {dN \over dy\,dp_{\rm T}^2\, d\phi}}
   \,.
 \eeq
In symmetric collision systems (which are the only ones we consider
here) the odd order coefficients $v_1,v_3,\dots$ vanish at midrapidity 
$y{\,=\,}0$ by symmetry. We here concentrate on the second harmonic 
coefficient which is conventionally called ``elliptic flow''. The 
$v_i$ are only defined at freeze-out but we already discussed how 
$v_2$ and $v_{2,p_{\rm T}^2}$ can be related to $\epsilon_p$ which 
is known also before freeze-out. 

$\epsilon_x$ and $\epsilon_p$ are functions of time; in the present 
section, however, we only need the {\em initial} spatial deformation 
$\epsilon_x(\tau_0)$ and the {\em final} momentum-space deformation 
$\epsilon_p(\tau_{\rm f})$. For simplicity we will quote them as 
$\epsilon_x$ and $\epsilon_p$, respectively, without the time 
arguments.

\suse{$p_{\rm T}$-dependence of elliptic flow}
\label{sec4a}

\begin{figure}[htbp]
 \hspace*{-0.3cm}\epsfig{file=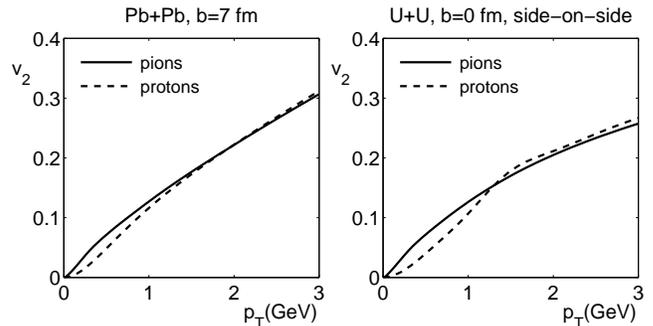,width=8.4cm}
 \caption{$p_{_{\rm T}}$-dependence of the elliptic flow coefficient
     $v_2$ for pions (solid) and protons (dashed), for 158 $A$\,GeV/$c$
     Pb+Pb collisions at $b{\,=\,}7$\,fm (left panel) and 
     155 $A$\,GeV/$c$ U+U collisions at $b{\,=\,}0$ in the side-on-side 
     configuration (right panel).
 \label{F12}}
 \end{figure}

Since most experiments have a limited acceptance in transverse 
momentum, the measured elliptic flow signal must be corrected for 
the $p_{_{\rm T}}$-acceptance. In Fig.~\ref{F12} we show the 
$p_{_{\rm T}}$-dependence of $v_2$ for pions and protons for 
semiperipheral Pb+Pb and central U+U collisions. In spite of 
their different masses, the predicted $v_2(p_{_{\rm T}})$ is 
rather similar for the two particle species \cite{HL99}. At low
$p_{_{\rm T}}$, the heavier protons show even a little less 
elliptic flow than the pions. To the extent that hydrodynamics 
is applicable, the larger $\la v_2\ra$ for protons than pions 
measured by NA49 \cite{NA49flow} is thus predominantly due to 
the different $p_{_{\rm T}}$-windows for the two particle 
species (the proton elliptic flow was measured at higher 
$p_{_{\rm T}}$ \cite{NA49flow}).

According to general arguments \cite{D95}, $v_2$ must vanish with 
zero slope as $p_{_{\rm T}}\to 0$. We checked that this is true. 
Fig.~\ref{F12} shows, however, that for pions the turnover from a 
roughly linear behaviour at large $p_{_{\rm T}}$ to zero slope as 
$p_{_{\rm T}}\to 0$ occurs at very small $p_{_{\rm T}}$-values,
$p_{_{\rm T}}<0.1$\,GeV/$c$; for protons the corresponding scale is 
somewhat larger. We have no quantitative analytic understanding of 
this momentum scale but note that qualitatively similar behaviour was 
found in \cite{BS00} using the kinetic UrQMD model.

\suse{Impact parameter dependence of elliptic flow}
\label{sec4b}

As one changes the impact parameter, the initial spatial deformation
$\epsilon_x$ of the transverse cross section through the reaction 
zone varies as shown in Fig.~3 of Ref.~\cite{O92}. The stronger the 
initial ellipticity, the stronger is the hydrodynamic response to it, 
i.e. the larger are $v_2$ or $\epsilon_p$ at freeze-out. Ollitrault
\cite{O92} showed that for an EOS with a constant velocity of sound, 
$\P e/\P p$\,=\,const., the ratio $\epsilon_p/\epsilon_x$ or, 
equivalently, $v_2/\epsilon_x$ is independent of the impact parameter 
\cite{fn3}. (Ollitrault \cite{O92} used the variable $v_{2,p_{\rm T}^2}$ 
which is closely related to $\epsilon_p$ \cite{fn2}. For pions $v_2$ 
and $\epsilon_p$ are related by a factor 2 \cite{KSH99}.) This scaling 
is broken only for very peripheral collisions which freeze out before 
the elliptic flow builds up and saturates; thus in hydrodynamics 
$v_2/\epsilon_x$ is constant over most of the impact parameter range.

A phase transition is characterized by a strong drop of the sound 
velocity in the critical region (for a first order phase transition 
the sound velocity vanishes in the mixed phase). It is therefore 
interesting to reinvestigate the impact parameter dependence of 
$v_2/\epsilon_x$ in the presence of a phase transition. The impact 
parameter not only controls the initial spatial ellipticity of the 
fireball, but also (with less variation) its initial energy density.
At a given beam energy, it is therefore possible to probe the EOS 
over a range of energy densities by varying the impact parameter. For
a beam energy, at which in central collisions the initial energy density
is not too far above the phase transition, it may thus be possible to
study the effect of the reduced sound velocity near the phase transition 
on the elliptic flow by changing the impact parameter. Weak structures
in Fig.~9 of Ref.~\cite{O92} first indicated that the quark-hadron
phase transition might thus become visible. Our analysis improves on
that analysis by including resonance decays which tend to dilute the
elliptic flow signature \cite{KSH99}.

\begin{figure}[htbp]
 \epsfig{file=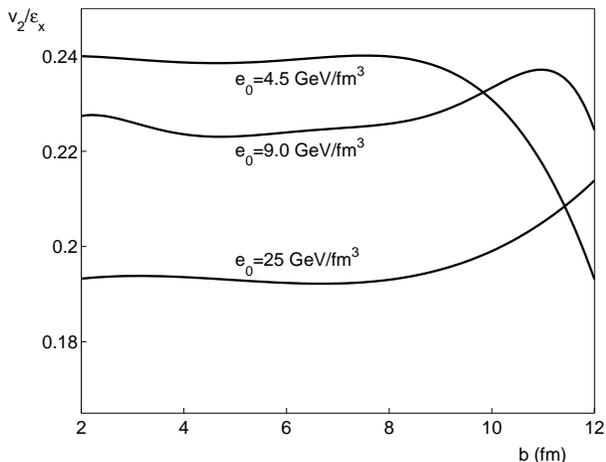,width=8cm}
 \caption{The ratio of the elliptic flow coefficient $v_2$ and
    the initial spatial ellipticity $\epsilon_x$ as a function
    of impact parameter $b$ for Pb+Pb collisions. Results for 
    three values of the initial central energy density at 
    $b{\,=\,}0$ ($e_0$\,=\,4.5, 9.0 and 25 GeV/fm$^3$) are shown.
    Note the suppressed zero on the vertical axis. 
 \label{F13}}
 \end{figure}

In Fig.~\ref{F13} we study the impact parameter dependence of 
$v_2/\epsilon_x$ in Pb+Pb collisions for three different initial 
central energy densities: $e_0{\,=\,}25$ GeV/fm$^3$ (corresponding
to a low energy run at RHIC), $e_0{\,=\,}9$ GeV/fm$^3$ (corresponding
to collisions at the highest SPS energy of 158\,$A$\,GeV), and
$e_0{\,=\,}4.5$ GeV/fm$^3$ (corresponding to lower SPS energies 
around 40\,$A$\,GeV). The calculated total pion multiplicity densities
at $b{\,=\,}0$ and midrapidity are 
${dN_\pi\over dy}\big\vert_{y=0}(b{\,=\,}0)$\,=\,859, 460, and 317, 
respectively. For later comparison with Fig.~\ref{F14} we also 
quote the corresponding rapidity densities for semiperipheral Pb+Pb
collisions: ${dN_\pi\over dy}\big\vert_{y=0}(b{\,=\,}7\,{\rm fm})$ 
=\,415, 220, and 148, respectively. 

Fig.~\ref{F13} shows that, at moderate impact parameters, the 
largest elliptic flow is generated at the lowest of these three
beam energies. At very large impact parameters (where hydrodynamics
becomes doubtful) the generated elliptic flow naturally drops to 
zero, since the overlap region and its initial energy density are 
then too small and the matter freezes out before flow can develop. 
What is interesting, however, is that at higher beam energies the 
elliptic flow starts out lower than at $e_0{\,=\,}4.5$\,GeV/fm$^3$, 
but then $v_2/\epsilon_x$ {\em rises} with increasing $b$. In fact, 
for $e_0{\,=\,}9$\,GeV/fm$^3$ this ratio reaches at $b{\,=\,}11$\,fm 
nearly the same value as for central collisions at 
$e_0{\,=\,}4.5$\,GeV/fm$^3$.

The decrease with rising beam energy of $v_2/\epsilon_x$ at moderate 
impact parameters was found \cite{KSH99} to result from the softening 
of the EOS in the phase transition region. The soft matter near the 
transition point inhibits the buildup of elliptic flow. Going at 
fixed beam energy to larger impact parameters is like going at 
fixed impact parameter to lower beam energies: in both cases the
initial energy density in the collision zone is reduced, and 
eventually the matter is dominated again by the relatively hard
hadron gas. When read from right to left, the curves in Fig.~\ref{F13} 
can thus be viewed as different projections of the excitation 
function of elliptic flow which will be discussed below. We emphasize 
in particular the rise of $v_2/\epsilon_x$ towards larger impact 
parameters at the high SPS and the low RHIC energy: without a phase 
transition this feature would be absent. Unfortunately, these 
variations are small (at the level of a few percent), and very 
accurate measurements are required to identify them.

Preliminary data from 158\,$A$\,GeV Pb+Pb collisions \cite{NA49flow}
show a monotonous decrease of $v_2/\epsilon_x$ with increasing impact 
parameter, instead of the nearly constant behaviour predicted by 
hydrodynamics (see Fig.~\ref{F13}). For $b\to 0$, however, the data 
seem to approach the hydrodynamic prediction. It is possible that 
semiperipheral Pb+Pb collisions do not equilibrate quickly enough to 
permit the elliptic flow to fully reach the hydrodynamic limit. Indeed, 
kinetic simulations with the RQMD code \cite{S99,NA49flow,VP00},
where the collision centrality is coupled to the degree of local
thermalization, are able to qualitatively explain the observed
decrease of $v_2/\epsilon_x$ with increasing impact parameter: more
peripheral collisions lead to less equilibration and hence to a weaker
elliptic flow response to the initial spatial ellipticity. When RQMD
is modified to simulate an EOS with a quark-hadron phase transition
\cite{S99}, the same generic decrease is superimposed on the rise of
$v_2/\epsilon_x$ at large $b$ shown here (middle curve in
Fig.~\ref{F13}); this results in a decrease of $(v_2/\epsilon_x)(b)$
which is first steep, then flattens, then finally steepens again
\cite{S99}. 

It is evident that a proper understanding of the interesting features
in the impact parameter dependence of $v_2/\epsilon_x$ predicted
in \cite{S99} for Pb+Pb collisions require the separation of 
pre-equilibrium effects from those induced by the softening of the 
EOS near the phase-transition. A collision system which is large 
enough to ensure sufficiently rapid thermalization for hydrodynamics
to apply would make life much easier. We therefore suggest to study 
elliptic flow in side-on-side U+U collisions at zero impact parameter
and search for the hydrodynamically predicted phase transition
signatures in the beam energy dependence of elliptic flow.  

\suse{Beam energy dependence of elliptic flow}
\label{sec4c}

The time-dependence of the flow patterns discussed in Sec.\,\ref{sec3} 
reflects itself also in the beam energy dependence of elliptic
flow. We already noted in \cite{KSH99} that the phase transition
causes an non-monotonic excitation function for the elliptic flow
coefficient $v_2$: as the collision energy is increased,
$v_2$ first rises (at low energies the fireball freezes out before
the elliptic flow can saturate) but then decreases again as the 
initial energy density rises above the QGP threshold. We now
understand that this decrease is intimately connected to the diluting
effects of the selfsimilar fireball expansion in the mixed phase,
even before the spatial deformation $\epsilon_x$ passes through zero
(see the discussion in Sec.\,\ref{sec3c}.) Without a phase transition
(EOS~H) this does not happen (see dash-dotted lines in
Fig.\,\ref{F14}); the slight decrease of $v_2$ with EOS~H at
asymptotically high energies has a different origin, namely a
reduction of $\epsilon_p$ by the opposite sign of the spatial
fireball anisotropy after $\epsilon_x$ has passed through zero.

 \begin{figure}[htbp]
 \hspace*{-0.2cm}\epsfig{file=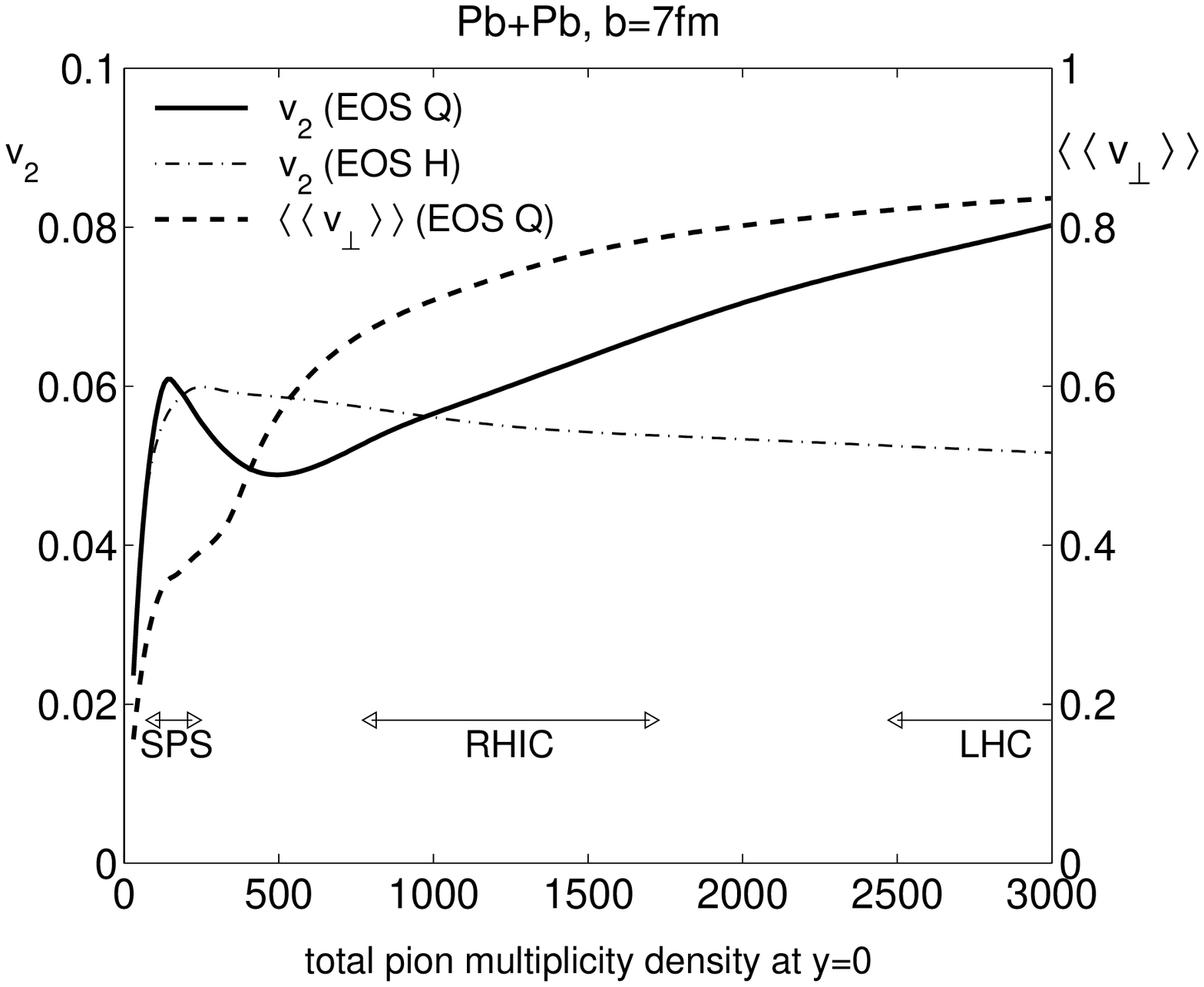,width=8.5cm}\\
 \hspace*{0.15cm}\epsfig{file=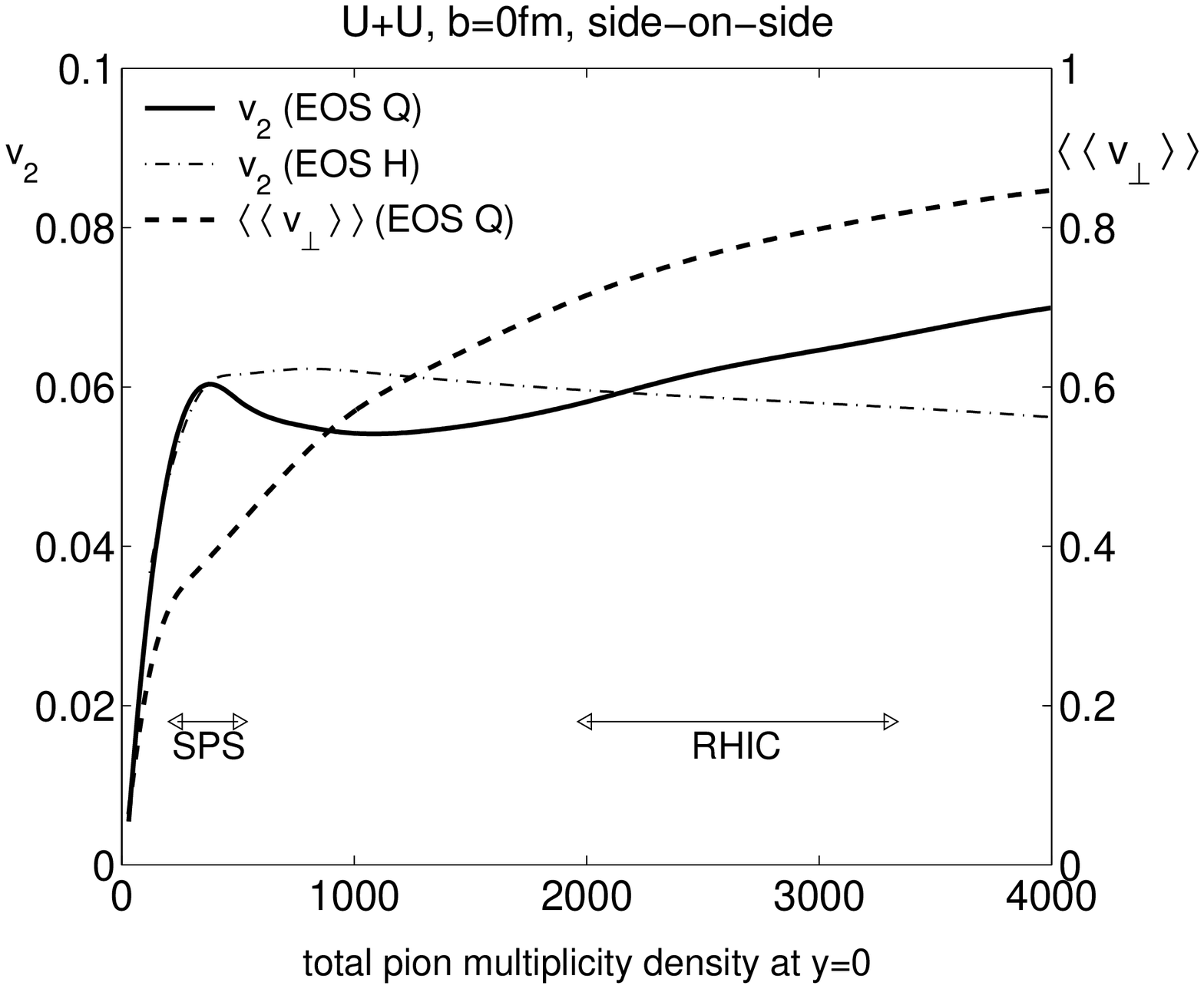,width=8.5cm}\\
 \caption{Excitation function of the elliptic flow coefficient $v_2$
   (left vertical axis) and the radial flow $\lda v_\perp \rda/c$ 
   (right vertical axis), for Pb+Pb collisions at $b{\,=\,}7$\,fm
   (upper panel) and side-on-side U+U collisions at $b{\,=\,}0$
   (lower panel). The horizontal axis gives the total pion
   multiplicity density at midrapidity, ${dN_\pi\over
     dy}\big\vert_{y=0}$, as a measure for the collision energy. 
   Horizontal arrows indicate the regions covered by SPS, RHIC, 
   and LHC. In the lower panel LHC would start around 5000. 
 \label{F14}}
 \end{figure}

The comparison of semiperipheral Pb+Pb collisions with central U+U
collisions in the upper and lower panels of Fig.\,\ref{F14} shows that 
this non-monotonic behaviour of the excitation function for $v_2$ is
not sensitive to the exi\-stence of the ``nutcracker phenomenon'': the
decrease of $v_2$ below its maximum in the SPS regime is only
slightly weaker in the U+U case than for Pb+Pb, although only the
latter features a ``cracking nut''. Since elliptic flow is a fragile
phenomenon which is quite sensitive to incomplete thermalization, we
believe that the most promising route towards experimental verification 
of the phase transition signature suggested here is to study the 
excitation function of $v_2$ in largest available deformed
collision system, namely central side-on-side U+U collisions. 
 
In Ref.\,\cite{KSH99} we missed the fact that at asymptotically high
energies the elliptic flow coefficient $v_2$ must approach the larger
value corresponding to the stiffer QGP equation of state EOS~I. We 
calculated in \cite{KSH99} the excitation function for $b{\,=\,}7$\,fm 
Pb+Pb collisions only up to multiplicity densities 
${dN_\pi\over dy}\big\vert_{y=0}{\,=\,}500$ and concluded prematurely 
that $v_2$ saturates at high collision energies at a value below the 
value corresponding to EOS~H. Fig.\,\ref{F14} extends the excitation 
functions for both Pb+Pb and U+U collisions to LHC energies and 
demonstrates that $v_2$ begins to rise again, eventually approaching 
the EOS~I limit. The dip, which indicates the presence of the phase 
transition, thus only covers the energy range between SPS and RHIC. Note 
that in the same energy region also the radial flow $\lda v_\perp\rda$ 
(dashed lines in Fig.\,\ref{F14}) is predicted to grow more slowly 
with $\sqrt{s}$ than at lower and higher beam energies where the 
expansion is predominantly driven by pure HG or pure QGP matter. 

\suse{Elliptic flow as an estimator for the thermalization time scale}
\label{sec4d}

Throughout this paper we have assumed early thermalization followed 
by hydrodynamic expansion. For a given initial deformation of 
the collision zone in the transverse plane (which can be calculated 
from geometry once the impact parameter is known, for example by a 
measurement of the number of spectator nucleons), this guarantees the 
maximum possible momentum-space response in the form of elliptic 
flow. Any delay in the thermalization process will lead to a reduction 
of the elliptic flow: even without secondary collisions the spatial
deformation of the region occupied by the produced particles decreases
by free-streaming, and if thermalization effectively sets in later,
the resulting anisotropies in the pressure gradients will be smaller,
leading to less elliptic flow.

We can use the above demonstrated fact that, up to variations of the 
order of 20\%, the hydrodynamic response $v_2$ to the elliptic spatial 
deformation at thermalization is essentially constant: 
$v_2^{\rm hydro}/\epsilon_x\approx$\,const.\,$\approx 0.25$. This allows 
to interpret the measured $v_2$ in terms of an {\em effective} initial 
spatial deformation at the point of thermalization, i.e. at the beginning of 
the hydrodynamic evolution. It is clearly not a good approximation to 
idealize the initial kinetic equilibration stage of the collision by 
a stage of collisionless free-streaming followed by hydrodynamic 
expansion, thereby assuming a sudden, but delayed transition from 
a non-equilibrium initial state to a fully thermalized fluid. Still,
this simple-minded picture can be used to obtain a rough first 
order-of-magnitude guess of the thermalization time scale, based on 
a measurement of $v_2$.

To this end we note that under free-streaming the phase-space 
distribution evolves as
 \beq{freestr}
   f(\bbox{r},\bbox{p},t) = 
   f\left(\bbox{r}-{\bbox{p}\over E}(t-t_0),\bbox{p},t_0\right)\,.
 \eeq
Using a Gaussian parametrization for the initial phase-space distribution
of produced secondary particles,
 \beq{Gauss}
   f(\bbox{r},\bbox{p},t_0) = 
   \exp\left[ -{x^2\over 2R_x^2}-{y^2\over 2R_y^2}
              -{p_x^2+p_y^2\over2\Delta^2}\right]\,,
 \eeq
one easily finds
 \bea{eps}
  \epsilon_x(t) &=& 
  {\int d^2r\,r^2\cos(2\phi_r) \int d^3p f(\bbox{r},\bbox{p},t)
   \over
   \int d^2r\,r^2 \int d^3p f(\bbox{r},\bbox{p},t)}
 \nonumber\\
  &\approx& \epsilon_x(t_0)\, {R_x^2+R_y^2
                               \over R_x^2+R_y^2 + 2 (c\Delta t)^2}\,,
 \eea
where $\Delta t = t{-}t_0$ is the time delay between particle formation 
and thermalization. Assuming that $\epsilon_x(t_0{+}\Delta t)$ can be 
obtained from the measured $v_2$ by dividing by $\approx\,0.25$, we can 
extract $\Delta t$ by rewriting 
(\ref{eps}) as
 \beq{ratioeps}
   {\epsilon_x(t_0+\Delta t)\over \epsilon_x(t_0)} =
   \left[ 1 + {(c\Delta t)^2 \over R^2 (1+\delta^2)}\right]^{-1}\,,
 \eeq
where $\delta$ parametrizes the initial deformation via $R_x=R(1{-}\delta)$,
$R_y=R(1{+}\delta)$ such that $\epsilon_x(t_0) = 2\delta/(1{+}\delta^2)$.

Inserting appropriate values for $R$ and $\delta$ one finds that for
Pb+Pb collisions at $b{\,=\,}7$\,fm a dilution by 50\% of the elliptic flow 
signal by initial free-streaming requires a time-delay of order 3.5 fm/$c$
until thermalization sets in; for central U+U collisions in the 
side-on-side configuration $\Delta t\,\approx\,5$\,fm/$c$ of approximate
free-streaming would be required to dilute the elliptic flow signal by 50\%.
This (admittedly rough) exercise demonstrates two points: (i) U+U 
collisions provide the better chance to observe the full hydrodynamic 
elliptic flow signal, and (ii) the observation of less elliptic flow than
hydrodynamically expected can be used to obtain a rough estimate of 
the thermalization time scale in the initial collision stage.

\se{Summary}
\label{sec5}

On the basis of hydrodynamic simulations we analyzed the sensitivity of
radial and elliptic transverse flow at midrapidity to the quark-hadron 
phase transition. We modelled this phase transition as a strongly first 
order phase transition with a latent heat of about 1.15 GeV/fm$^3$. 
It manifests itself dynamically as an expan\-ding shell of mixed phase 
matter inside which all pressure gradients and thus all hydrodynamic
acceleration forces vanish. Compared to the situation of a pure HG or
a pure QGP phase this leads to a reduction of both radial and elliptic 
flow. Elliptic flow, as the more fragile phenomenon which is generated
only by azimuthal anisotropies in the pressure gradients, shows a larger
sensitivity to the phase transition than radial flow. Also, since we 
showed that it saturates well before freeze-out, it more directly 
reflects the EOS during the early and dense stages of the expansion. 

As a tell-tale signature for the phase transition we predict a 
non-monotonic excitation function for the elliptic flow coefficient 
$v_2$ as shown in Fig.~\ref{F14}. In the present paper we explored
in great detail the origin of the dip in $v_2$, which we predict 
to occur in the energy region between the SPS and RHIC, by performing 
a careful ana\-ly\-sis of the space-time evolution of the anisotropic 
transverse flow pattern for a variety of collision energies. As the 
dynamical origin of the phase transition signature in $v_2$ we 
identified the existence of a large subvolume of mixed phase matter 
which undergoes nearly selfsimilar, acceleration-free expansion while 
it lasts. In addition to the $v_2$ excitation function it leaves 
traces in the impact parameter dependence of the response 
$v_2/\epsilon_x$ of the elliptic flow to the initial spatial 
deformation of the collision zone, and in the (not directly 
measurable) time evolution of the flow anisotropy $\epsilon_p$.

When colliding spherical nuclei with each other, the measurement of
elliptic flow requires selecting collisions at rather large impact
parameters ($b\gtrsim5$\,fm) in order to achieve a sufficiently large
initial spatial deformation of the nuclear overlap region (reaction 
zone). Correspondingly the overall size of the elliptically deformed, 
expanding fireball is small, and one may doubt the applicability of 
our hydrodynamic approach. We here point out that central U+U 
collisions in the side-on-side configuration provide nearly twice 
larger collision volumes at similar deformation as Pb+Pb collisions 
at $b{\,=\,}7$\,fm and should thus exhibit hydrodynamic behaviour 
much more clearly.

We therefore carefully compared central side-on-side U+U collisions 
with semipe\-ri\-phe\-ral Pb+Pb collisions at all collision 
energies. We showed that the phase transition signature in the $v_2$
excitation function manifests itself similarly in both
collision systems. The U+U system should thus be preferred for 
its presumed better hydrodynamical behaviour and for the larger
particle multiplicities which improve the statistics of elliptic flow
measurements. The phase transition signal appears to be slightly 
stronger in the smaller Pb+Pb system; we were able to trace this to
the ``nutcracker phenomenon'' of Shuryak and Teaney \cite{TS99} which,
unfortunately, only occurs in the Pb+Pb system. In trying to understand 
the fragility of ``nutcracker flow'' we found that it crucially relies
on the existence of a rather thick shell of mixed phase matter
{\em at rest} in the initial state of fireball expansion, which 
surrounds a significant core of QGP. In response to internal 
pressure gradients the QGP core starts to expand and ``slams'' into 
the surrounding shell of mixed phase at rest. This cannot happen in
central U+U collisions since there the initial transverse energy 
density profile drops to zero so steeply that no visible mixed phase 
shell forms.

We thus conclude that the interesting ``nutcracker flow'' phenomenon 
constitutes a very fragile variant of anisotropic flow which is not 
generated in central U+U collisions. If the fireballs formed in 
semiperipheral Pb+Pb collision should turn out to be too small to 
achieve sufficient local thermalization for hydrodynamics to work, it
may be unmeasurable. Fortunately, the elliptic flow signature for
the phase transition is more robust and does not require the actual
``cracking of the nut''; it should be clearly visible in central U+U
collisions.

This raises the question how to experimentally select the side-on-side 
collision geometry. By requiring zero spectators one can trigger on 
configurations in which the colliding nuclei overlap completely in the
transverse plane. This still allows for arbitrary, but (up to a sign) 
equal angles ($\theta_1=\pm \theta_2$) between the beam direction and 
the long axes of the two deformed nuclei. The interesting side-on-side 
configuration corresponds to $\theta_1{\,=\,}\theta_2{\,=\,}90^\circ$. 
Since this configuration has the largest initial spatial deformation 
in the transverse plane, it generates the largest elliptic flow $v_2$; 
therefore, Shuryak \cite{Sh00} suggested a cut on large $v_2$ to select 
the side-on-side collision geometry. Unfortunately, the event-by-event 
fluctuations of $v_2$ are so large that this off-line trigger is not 
expected to be very efficient \cite{Posk}; furthermore, it would 
introduce an inconvenient trigger bias into our suggested investigation
of the dependence of $v_2$ on various control parameters.    

We have not been able to come up with a more efficient selection 
criterium. We checked that with initial conditions calculated according 
to (\ref{init}), the produced charged particle multiplicity densities at
midrapidity vary by less than 5\% between tip-on-tip and side-on-side
collisions (with side-on-side collisions producing more particles, with 
slightly smaller $\langle p_T\rangle$ at freeze-out). Again this 
difference is well below the expected level of event-by-event 
fluctuations. Its smallness is explained by the fact that with the 
{\em Ansatz} (\ref{init}) the amount of entropy $dS/dy$ stopped at 
midrapidity is essentially independent of the orientation 
$\theta_1=\pm\theta_2$ (for 0$^\circ$ it is 1.3\% larger than for 
90$^\circ$), and boost-invariant longitudinal expansion conserves 
$dS/dy$. At higher collision energies minijet production may overtake
the soft particle production processes implicitly assumed in (\ref{init});
instead of scaling with the number of wounded nucleons as in (\ref{init}),
minijet production scales with the number of nucleon-nucleon collisions,
involving the product rather than the sum of the nuclear thickness 
functions appearing in (\ref{init}). In this case tip-on-tip collisions
are expected to generate considerably more entropy in the transverse
plane at midrapidity than side-on-side collisions, and one could trigger
on the latter by selecting for zero spectators combined with low 
$dN/dy(y{=}0)$. --- In the absence of an efficient trigger for side-on-side
U+U collisions at present-day collision energies one will be forced 
to compare with data which are averaged over all orientations 
$\theta_1=\pm\theta_2$. The computation of an orientation-averaged 
excitation function for $v_2$ is, however, numerically expensive;
we therefore postpone it until experiments involving U+U are approved. 

Our prediction of a dip in the excitation function of $v_2$ at 
midrapidity is directly related to the one by Rischke {\it et al.} 
\cite{Rischke95} of a dip in the excitation function for directed 
flow at forward and backward rapidities: both rely on the softening 
of the EOS near the phase transition which results in reduced 
hydrodynamic pressure gradients. We point out, however, that, as 
the collision energy increases, the time interval during which 
directed flow is generated (the nuclear transition time) becomes 
shorter and shorter, and the prospects for sufficiently fast local 
thermalization to validate hydrodynamic concepts thus become
{\em worse and worse}. The opposite is true for elliptic flow: 
Figs.~\ref{F7}  and \ref{F9} show that the time interval over 
which elliptic flow builds up approaches at high collision energies 
a finite limit of about 7\,fm/$c$ for semiperipharal Pb+Pb and about
12\,fm/$c$ for central U+U collisions. The density of produced 
particles, on the other hand, continues to increase, leading to 
shorter and shorter thermalization times. The hydrodynamic 
description of elliptic flow buildup should thus become {\em better} 
with increasing collision energy. 
 
We finally comment on the sensitivity of the proposed phase
transition signature to our simple modelling of the phase transition: 
we used a Maxwell construction between the HG and QGP equations of state, 
leading to a strong first order phase transition with large latent 
heat. We don't believe that smoothing the phase transition to a rapid 
crossover will qualitatively alter our results: the only major change 
will be a replacement of the acceleration-free mixed phase by a 
transition region with non-zero, but nevertheless small pressure 
gradients. However, since elliptic flow signals are generically weak
and the predicted effects from the phase transition are at a level
of only about 10\% of this signal, further hydrodynamic simulations 
using a more realistic modelling of the EOS may be required for a 
reliable quantitative assessment of the expected experimental signal.
   
\acknowledgments

We gratefully acknowledge fruitful discussions with Tamas Bir\'o, 
Pasi Huovinen, Art Poskanzer, and Ser\-gei Voloshin. Our hydrodynamic 
code is a (2+1)-dimensional generalization of a (1+1)-dimensional 
algorithm for central collisions which was originally developed by 
M.~Kataja, P.V.~Ruuskanen, R.~Venugopalan, and P.~Huovinen. We thank 
these colleagues for allowing us to modify their code for non-central 
collisions. This work was supported in part by BMBF, DFG and GSI.

\appendix
\se{Implementation of boost invariance}
\label{appa}

An elegant method of introducing longitudinal boost invariance with 
the longitudinal velocity field $v_z=z/t$ makes use of the notation 
of general covariant derivatives. 

In an arbitrary coordinate system the equations of motion can be 
written as 
 \beq{A1}
   {T^{mn}}_{;m} = 0\,, \qquad 
   {j^m}_{;m} = 0\,,
 \eeq
where the semicolon indicates a covariant derivative. For tensors of 
rank 1 and 2 it reads explicitly
 \bea{A2}
   j^i_{;p} &=& j^i_{,p} + \Gamma^i_{pk}\, j^k \,,
 \\
 \label{A3}
   {T^{ik}}_{;p} &=& {T^{ik}}_{,p} + \Gamma^i_{pm} T^{mk} 
                                   + \Gamma^k_{pm} T^{im}\,,
 \eea
where the komma denotes a simple partial derivative and the Christoffel 
symbols $\Gamma^s_{ij}$ are given by derivatives of the metric tensor
$g^{ab}(x)$:
 \beq{A4}
    \Gamma^s_{ij} =
    \half g^{ks} \bigl( g_{ik,j} + g_{jk,i} - g_{ij,k}\bigr)\,.
 \eeq
We use this with the following transformation from Cartesian to light 
cone coordinates:
 \bea{A5}
   x^\mu=(t,x,y,z) & \longrightarrow  & \bar{x}^m=(\tau,x,y,\eta)
 \nonumber\\
   t = \tau \cosh\eta &     &  \tau = \sqrt{t^2-z^2}
 \\
   z = \tau \sinh\eta &     &  \eta = \half \ln{t{+}z\over t{-}z} \,.
 \eea
In the new coordinate system the velocity field (after inserting $v_z=z/t$)
is given by
 \beq{A6}
   \bar{u}^m = \bar{\gamma}(1,\bar{v}_x,\bar{v}_y,0)
 \eeq
with $\bar{v}_i \equiv v_i\cosh\eta$, $i=x,y$, and
$\bar{\gamma} \equiv 1/{\textstyle{\sqrt{1{-}\bar{v}_x^2{-}\bar{v}_y^2}}}$.

Now we turn to the metric of the new system. We have 
 \bea{A7}
   ds^2 = g_{\mu \nu} dx^\mu dx^\nu 
       &=& dt^2 - dx^2 - dy^2 - dz^2
  \nonumber \\
       &=& d\tau ^2 - dx^2 - dy^2 - \tau^2 d\eta^2
 \eea
and therefore
 \beq{A8}
   g_{mn}=\left( \begin{array}{*{4}{c}} 
        1 & 0 & 0 & 0 \\ 
        0 & -1 & 0 & 0 \\
        0 & 0 & -1 & 0 \\ 
        0 & 0 & 0 & -\tau^2 \\
        \end{array} \right)\,,
 \eeq
The only non-vanishing Christoffel symbols are
 \beq{A9}
   \Gamma^{\eta}_{\eta \tau} = 
   \Gamma^{\eta}_{\tau \eta} = {1\over\tau}\,,\qquad
   \Gamma^\tau_{\eta \eta} = \tau\,.
 \eeq
Finally, by making use of the relations
$T^{\tau i}= \bar{v}_i T^{\tau\tau}+\bar{v}_i p$ and 
$T^{\eta \eta}= p /\tau^2$ the energy-momentum conservation equations
(\ref{A1}) turn for $n=\tau,x,y,\eta$ into
 \begin{mathletters}
 \label{A10}
 \bea{A10a}
 \FL
   &&{T^{\tau\tau}}_{,\tau} + (\bar{v}_x T^{\tau\tau})_{,x}
     + (\bar{v}_y T^{\tau\tau})_{,y} =
 \\
   &&\qquad = -{p{+}T^{\tau\tau}\over \tau} - (p\, \bar{v}_x)_{,x}
              - (p\, \bar{v}_y)_{,y} \,,
 \nonumber\\
 \label{A10b}
   &&{T^{\tau x}}_{,\tau} + (\bar{v}_x T^{\tau x})_{,x}  
     + (\bar{v}_y T^{\tau x})_{,y} =
 \\
   &&\qquad = - p_{,x} - {T^{\tau x}\over \tau}\,,
 \nonumber\\
 \label{A10c}
   &&{T^{\tau y}}_{,\tau} + (\bar{v}_x T^{\tau y})_{,x}
     + (\bar{v}_y T^{\tau y})_{,y} =
 \\
   &&\qquad = - p_{,y} - {T^{\tau y}\over \tau}\,,
 \nonumber\\
 \label{A10d}
   &&{1\over \tau^2}\, p_{,\eta} = 0\,,
 \eea
 \end{mathletters}
while the current conservation (\ref{A1}) becomes
 \beq{A10e}
   {j^\tau}_{,\tau} + (\bar{v}_x j^\tau)_{,x} + (\bar{v}_y j^\tau)_{,y}
   = - {j^\tau\over \tau}\,.
 \eeq
We note the explicit appearance of $\tau$ on the r.h.s. of the 
differential equations, reflecting the dilution of the matter
due to the boost-variant longitudinal expansion. Connected with 
this is the initial equilibration time $\tau_0$ as one of the 
model parameters. Equation (\ref{A10d}) expresses the fact that, 
due to longitudinal boost-invariance, the evolution is 
$\eta$-independent.

Multiplying these equations by $\tau$ and introducing the scaled quantities
$\tilde{\jmath}^\mu=\tau j^\mu$, $\tilde{T}^{\mu\nu}=\tau T^{\mu \nu}$, and
$\tilde{p}=\tau\,p$ leads to the simple form (\ref{DGL}).



\begin{thebibliography}{99}
\bibitem{fn1}
  Note that the box distribution radii used in \cite{Sh00} are 4.6\% 
  larger than the Woods-Saxon radius parameters used by us. The values 
  for $R_{l,s}$ given in \cite{Sh00} were scaled down by this factor.
\bibitem{fn1a}
  For an approximately Gaussian initial transverse energy density profile
  $e_0{\,=\,}9$\,GeV/fm$^3$ at $\tau_0{\,=\,}0.8$\,fm/$c$ corresponds to 
  an average energy density $\bar e(0.8$\,fm$/c)=$ 4.5\,GeV/fm$^3$.
  Due to longitudinal Bjorken expansion this decreases with $\tau^{-4/3}$
  to $\bar e(1$\,fm$/c){\,=\,}3.34$\,GeV/fm$^3$ at $\tau{\,=\,}1$\,fm/$c$.
  This value is consistent with $\bar e(1$\,fm/$c)=$ $(3.2\pm0.3)$\,GeV/fm$^3$ 
  as estimated from the measured transverse energy $dE_T/dy$ using 
  Bjorken's formula \cite{e0}.
\bibitem{fn2}
  According to (\ref{CF}) the hadronic spectra from which $v_2$ and
  $v_{2,p_{\rm T}^2}$ are calculated involve an integration over the
  freeze-out surface $\Sigma$ whereas $\epsilon_p$ is given by a 
  spatial integral at fixed time (e.g. at $\tau_{\rm f}$). We found 
  that the difference is small since most particles freeze out during 
  a very short time interval. This is in particular true at very high 
  energies (see Fig.~4 in the first paper of Ref.~\cite{KSH99}). Another 
  difference between $\epsilon_p$ and $v_{2,p_{\rm T}^2}$ is that the 
  first implicitly sums the contributions from all particle species while
  the latter is defined in terms of the spectrum of a single particle
  species (in our case pions). The difference is small for 
  pion-dominated systems like those studied here. For the same 
  reason, however, the $v_{2,p_{\rm T}^2}$ for protons or kaons 
  may well differ from $\epsilon_p$.
\bibitem{fn3}
  The differing statement in \cite{HL99} that this ratio should 
  increase with decreasing impact parameter (see Fig.~5 in \cite{HL99})
  is incorrect. It is derived from schematic considerations which 
  assume that elliptic flow scales like radial flow, not taking into
  account the earlier saturation of the former.
\end{thebibliography}
\end{document}